\documentclass[aps,prd,10pt,twocolumn,nofootinbib,groupedaddress,amsfonts,floatfix]{revtex4-1}

\usepackage[colorlinks=true,urlcolor=blue,linkcolor=,citecolor=blue]{hyperref}
\usepackage{amsmath,amssymb,amstext,amsbsy,amsfonts,amsthm,graphicx,microtype,dsfont}
\usepackage[capitalize]{cleveref}

\newcommand{\Tr}[1]{\text{Tr}\left[{#1}\right]}
\creflabelformat{equation}{(#2\textup{#1}#3)}
\newcommand{\abs}[1]{{\left \vert #1 \right \vert}}

\usepackage{color}
\usepackage{ifthen}
\newboolean{editorial}
\setboolean{editorial}{true}
\newcommand{\editorial}[2]{\ifthenelse{\boolean{editorial}}{\textcolor{red}{[\textsf{\textbf{{#1}}}: }\textcolor{blue}{\textsf{{#2}}}\textcolor{red}{]}}{}}

\usepackage{xcolor}

\begin{document}

\title{Non-Abelian gauge preheating}

\author{Peter Adshead${}^1$}
\author{John T. Giblin, Jr${}^{2,3}$}
\author{Zachary J. Weiner${}^{1,2}$}
\affiliation{${}^1$Department of Physics, University of Illinois at Urbana-Champaign, Urbana, Illinois 61801, USA}
\affiliation{${}^2$Department of Physics, Kenyon College, Gambier, Ohio 43022, USA}
\affiliation{${}^3$CERCA/ISO, Department of Physics, Case Western Reserve University, Cleveland, Ohio 44106, USA}


\begin{abstract}
We study preheating in models where a scalar inflaton is directly coupled to a non-Abelian $SU(2)$ gauge field.
In particular, we examine $m^2 \phi^2$ inflation with a conformal, dilatonlike coupling to the non-Abelian sector.
We describe a numerical scheme that combines lattice gauge theory with standard finite difference methods applied to the scalar field.
We show that a significant tachyonic instability allows for efficient preheating, which is parametrically suppressed by increasing the non-Abelian self-coupling.
Additionally, we comment on the technical implementation of the evolution scheme and setting initial conditions.
\end{abstract}

\maketitle


\section{Introduction}

Recent studies of preheating following inflation have pushed beyond scalar-scalar studies to examine theories with field content that more closely resembles that of the Standard Model.
The dynamics of gauge fields may be critical to a variety of phenomena in the early Universe, such as preheating, magnetogenesis, and even inflation itself.
Recent focus has been applied to massless, Abelian gauge fields coupled both dilatonically and axially to scalar or pseudoscalar inflatons \cite{Dufaux:2010cf,Deskins:2013lfx,Adshead:2015pva, Lozanov:2016pac, Adshead:2016iae,Giblin:2017wlo}.
Several studies have employed numerical lattice simulations in order to capture the full nonlinear preheating dynamics. 
Simulations of continuum Abelian, $U(1)$ fields have demonstrated stability with respect to the gauge-fixing condition without appealing to lattice techniques~\cite{Deskins:2013lfx,Adshead:2015pva,Adshead:2016iae,Giblin:2017wlo}.
However, numerical investigations of non-Abelian fields demand additional effort in order to remain on the gauge-constraint surface.

Extending the understanding of inflaton-gauge interactions to non-Abelian sectors is of particular importance, since much of the known Standard Model gauge group is non-Abelian.
Gauge fields likely play a key role in the early Universe in generating the matter-antimatter asymmetry, for example in electroweak baryogenesis and leptogenesis.
Additional interest in detailed studies of non-Abelian fields in the early Universe comes from a growing number of inflationary models that contain non-Abelian fields~\cite{Maleknejad:2011jw,Maleknejad:2011sq,Adshead:2012kp,Martinec:2012bv,Adshead:2016omu,Adshead:2017hnc,Caldwell:2017chz,Karciauskas:2011fp}.
The inherent nonlinearities of $SU(2)$ theories naturally call for numerical study; however, this same property potentially sources instability.
In particular, it is imperative that the gauge condition be well satisfied within the bounds of numerical truncation error throughout the entirety of the simulation; typically, numerical implementations of such fields employ a reformulation of the gauge fields as defined by lattice gauge theory~\cite{Wilson:1974sk}, the principal tool used to study non-Abelian theories in quantum chromodynamics (QCD).

This work follows a number of numerical studies of non-Abelian gauge theories that implement lattice gauge theory in real time, including studies of preheating~\cite{GarciaBellido:1999sv,Moore:2001zf,Rajantie:2000nj,Smit:2002yg,Skullerud:2003ki,Tranberg:2003gi,GarciaBellido:2003wd,DiazGil:2008tf,DiazGil:2007dy,Figueroa:2015rqa,Ema:2016kpf,Enqvist:2015sua,Repond:2016sol,Tranberg:2017lrx}, electroweak baryogenesis~\cite{Saffin:2011kn,Mou:2017atl,Mou:2017zwe}, Chern-Simons theory~\cite{Moore:1996wn,Moore:1996qs,Moore:1997cr,Moore:1997sn,Berruto:2000dp,Bodeker:1999gx,Olesen:2015baa,Figueroa:2017qmv}, sphalerons~\cite{Grigoriev:1989je,Ambjorn:1990wn,Ambjorn:1990pu,Moore:1998swa}, cosmic strings~\cite{Bevis:2006mj,Dufaux:2010cf,Daverio:2015nva,Hindmarsh:2016lhy,Hindmarsh:2017qff,Hindmarsh:2016dha}, and other studies of free Yang-Mills theory~\cite{Kurkela:2012tq,Kurkela:2012hp,Kurkela:2016mhu}.
Our work differs from prior non-Abelian preheating studies by considering direct couplings between an \textit{uncharged} scalar, which drives inflation, and the $SU(2)$ sector, rather than coupling the $SU(2)$ fields to a charged field such as the Higgs.
Scalar-gauge couplings of the form $f(\phi) F F$ or $f(\phi) F \tilde{F}$ often~\cite{Deskins:2013lfx,Adshead:2015pva,Adshead:2016iae} (though not universally~\cite{Giblin:2017wlo}) exhibit tachyonic instabilities leading to highly efficient production of gauge bosons.

In this work we study the model described by the Lagrangian density
\begin{equation}
	\mathcal{L} = \frac{m_\text{pl}^2}{16\pi} R - \frac{1}{2} \partial_\mu \phi \partial^\mu \phi - V(\phi) - \frac{ \Theta(\phi) }{2} \Tr{F_{\mu\nu} F^{\mu\nu}},
	\label{continuumaction}
\end{equation}
where $F_{\mu\nu}$ is the usual field strength of an $SU(2)$ gauge field, the trace is over Pauli matrices with normalization defined in \cref{paulidef}, and $\Theta(\phi)$ is a coupling function.
This model is analogous to the model studied in Ref.~\cite{Deskins:2013lfx}, except here we consider the gauge group $SU(2)$ rather than $U(1)$.
In the Abelian context, this model has been extensively studied as a source of magnetogenesis from inflation~\cite{Turner:1987bw,Ratra:1991bn,Lemoine:1995vj,Bamba:2003av,Martin:2007ue,Demozzi:2009fu,Kanno:2009ei,Subramanian:2009fu,Durrer:2013pga,BazrafshanMoghaddam:2017zgx} as the coupling function $\Theta(\phi)$ breaks the conformal invariance of the gauge fields, permitting the generation of primordial magnetic fields.
However, recent work has cast doubt on such models' efficacy for magnetogenesis due to the so-called strong coupling problem~\cite{Demozzi:2009fu}.
Here we consider the role such couplings might play after inflation to rapidly and efficiently drain the energy in the homogeneous inflaton condensate via the production of gauge bosons.

After inflation, the scalar field $\phi$ oscillates about the minimum of its potential, which we take to be quadratic,
\begin{align}
	V(\phi) = \frac{1}{2} m_\phi^2 \phi^2.
	\label{infpot}
\end{align}
We set the inflaton mass $m_\phi = 10^{-6} \, m_{\rm pl}$, consistent with the amplitude of the observed scalar spectrum~\cite{Planck:2013jfk}. 
The particular form of the dilatonic coupling we consider is
\begin{align}
	\Theta(\phi) = e^{\phi/M},
\end{align}
which is parametrized by a mass scale $M$ and tends to unity as the inflaton decays, recovering the canonical Yang-Mills action.
We extend the analysis of \cite{Deskins:2013lfx} by allowing the initial value of the scalar field to be positive \textit{or negative}.
In the former case, $\Theta(\phi) = \exp\left(+ \abs{\phi}/M\right)$ is large, which might spoil the negative pressure that facilitates accelerated expansion.
On the other hand if the field rolls in from negative values, $\phi <0$, then $\Theta(\phi) = \exp\left(- \abs{\phi}/M\right)$ serves as an exponential suppression of the gauge field contribution during inflation.

We work in conformal Friedmann-Lema\^itre-Robertson-Walker (FLRW) spacetime under the ``mostly plus'' sign convention,
\begin{align}
	ds^2 = a(\tau)^2 ( - d\tau^2 + d{\vec{x}}^2 ),
\end{align}
where primes denote derivatives with respect to conformal time.
The evolution of the scale factor $a(\tau)$ is given by the Friedmann equation,
\begin{align}
	\mathcal{H}(\tau)^2 &= \frac{ 8 \pi }{3 m_\text{pl}^2} T_{00}(\tau).
\end{align}
The total energy density is related to the stress-energy tensor via $T_{00}(\tau) = a(\tau)^2 \rho_\text{tot}(\tau)$, and has a contribution from the scalar field
\begin{align}
	\rho_\phi &= 
	\frac{1}{2 a^2} {\phi'}^2 + \frac{1}{2 a^2} (\nabla\phi)^2 + V(\phi)
\end{align}
as well as a contribution from the gauge field,
\begin{align}
	\rho_\text{gauge} = \frac{\Theta(\phi)}{2 a^4} \left[ \sum_{i=1}^3 \left( F_{0i}^a \right)^2 + \sum_{j > i} \left( F_{ij}^a \right)^2 \right].
\end{align}
The equation of motion for the inflaton is
\begin{align}\label{phiEOM}
	\phi'' = \nabla^2\phi - 2 \mathcal{H} \phi' - a^2 \frac{1}{4} \frac{d \Theta}{d \phi} F_{\mu\nu}^a F_a^{\mu\nu} - a^2 \frac{dV}{d\phi}.
\end{align}
Turning to the gauge sector, the non-Abelian field tensor is
\begin{align}
	F_{\mu\nu}^a &= \partial_\mu A_\nu^a-\partial_\nu A_\mu^a + g f^{abc}A_\mu^b A_\nu^c,
\end{align}
where the $SU(2)$ structure constants are given by the three-dimensional Levi-Civit\`a symbol, $f^{abc} = \epsilon^{abc}$, and $g$ is the gauge self-coupling constant.
Sums over repeated flavor indices are implied, regardless of their placement.

Simulations of gauge theories must first initialize a field configuration that satisfies the constraint from Gauss's law and a chosen gauge-fixing condition, and they must subsequently preserve the satisfaction of these constraints throughout the evolution.
We first discuss the methodology that allows us to satisfy the gauge constraint: lattice gauge theory. 


\section{The lattice approximation}

First introduced by Wilson in 1974~\cite{Wilson:1974sk} and ubiquitously implemented in studies of lattice QCD as well as most numerical studies of non-Abelian gauge fields,
lattice gauge theory recasts the gauge fields as fields of ``link variables,''
\begin{align}
	U_\mu(x) &= \exp\left( - a_\mu g A_\mu^a(x) \sigma^a \right),
\end{align}
which quantify the gauge connection between lattice sites.
In this expression, $a_\mu$, where $\mu \in \{0,1,2,3\}$, are the lattice spacings and $\sigma^a$ are the Pauli matrices [defined in \cref{paulidef}].
Note that when working with the lattice variables we drop the Einstein summation convention.
As the links belong to the group $SU(2)$, we choose to evolve the coefficients of the links' decomposition onto the Pauli basis
\begin{align}
	U_\mu(x) &= U_\mu^0 \cdot \mathds{1}_{2\times 2} + U_\mu^a \sigma^a.
\end{align}
The field strength is given by closed loops in the lattice called \textit{plaquettes}, defined by
\begin{align}
	P_{\mu\nu}(x) = U_\mu(x) U_\nu(x + \hat{\mu}) U_\mu^\dag(x + \hat{\nu}) U_\nu^\dag(x).
\end{align}
In this language, the lattice Lagrangian density is
\begin{align}
	\begin{split}
	\mathcal{L}_{G} &= \frac{2}{a_i^4 g^2} \frac{1}{a(t)^4} \Bigg( \frac{1}{\kappa^2} \sum_i \text{Tr}\left[1 - P_{0i}(x)\right] \\
	&\hphantom{={} \frac{2}{a_i^4 g^2} \frac{1}{a(t)^4} \Bigg(} - \sum_{j > i} \text{Tr}\left[1 - P_{ij}(x)\right] \Bigg),
	\end{split}
	\label{latticeaction}
\end{align}
where the lattice spacing ratio $\kappa = a_0 / a_i$.
This Lagrangian density (which defines the well-known Wilson action) is only equivalent to the continuum form, \cref{continuumaction}, in the limit of vanishing lattice spacing, $a_\mu \rightarrow 0$.
For finite lattice spacing, we have traded an evolution of the true continuum theory for the ability to natively and exactly satisfy the gauge condition.

Next, in these terms we may also produce the lattice form of the gauge energy density, separated into electric and magnetic components
\begin{align}
	\rho_E &= \frac{2 \Theta(\phi)}{a(t)^4 a_i^4 g^2} \frac{1}{\kappa^2} \sum_i \Tr{ 1 - P_{0i} },
\end{align}
and
\begin{align}
	\rho_B &= \frac{2 \Theta(\phi)}{a(t)^4 a_i^4 g^2} \sum_{i > j} \Tr{1 - P_{ij} }.
\end{align}

At this point we diverge from the traditional methods of lattice QCD, which typically employ Monte-Carlo methods in four-dimensional Euclidean space.
We instead aim to solve an initial value problem in $3+1$ FLRW spacetime with finite-difference methods.
While it is possible to discretize gauge theories in spatial dimensions while keeping time continuous~\cite{Kogut:1974ag}, here we retain the full $3+1$-dimensional discretization.
We work in temporal gauge, which sets the continuum fields $A_0^a = 0$; on the lattice this amounts to setting the temporal links $U_0(x)$ to the identity.
By varying the lattice action, \cref{latticeaction} (summed over lattice sites), with respect to the link coefficients $U_\mu^a$, one obtains equations of motion for the combination
\begin{align}
	E_i (x) = U_i(x + \hat{0}) U_i^\dag(x),
\end{align}
which we call the ``electric field.''
In turn, $E_i$ provides an ``update rule'' which relates the links at the next time step in terms of the current links and electric fields,
\begin{align}\label{linkupdate_text}
	U_i(x + \hat{0}) &= E_i(x) U_i(x).
\end{align}
The evolution of the electric field then follows:
\begin{align} 
\begin{split}
	E_i^a(x + \hat{0}) &= \frac{\Theta(x)}{\Theta(x + \hat{0})} E_i^a(x)
	- \kappa^2 \sum_{j\neq i} P_{ij}^a(x + \hat{0}) \\
	&\hphantom{={}} + \kappa^2 \sum_{j\neq i} \frac{\Theta(x + \hat{0} - \hat{\jmath})}{\Theta(x + \hat{0})} P_{ij}^a(x + \hat{0} - \hat{\jmath}).
\end{split}
\end{align}
Note that this equation of motion only holds for the trace-free part of the fields, i.e., the three ``Pauli flavors'' of the fields $E^a$, $a=1$, 2, or $3$.
The ``zero flavor'' of the electric fields is obtained via the unitarity the lattice variables: as elements of $SU(2)$, both the links and electric fields must satisfy
\begin{align}
	U_\mu^\dag U_\mu = E_\mu^\dag E_\mu = 1.
\end{align}
The unitarity of the link variables is automatically satisfied by the link update, \cref{linkupdate_text}, while this constraint must be used to obtain $E^0_i$ at each time slice.
See \cref{linksEOMappendix} for further exposition on the evolution of the lattice variables.

These update rules manifestly preserve the gauge condition throughout the evolution by exactly defining the field configurations on subsequent time slices.
That is, whereas numerical integration typically approximates the evolution equations describing the exact fields of the continuum theory, the lattice formalism instead places this approximation at the level of the theory itself.
In our $3+1$-dimensional context, we evolve a 3D field configuration which approximates its continuum form, the evolution itself being an approximation of the true continuum equations of motion.
In turn, the lattice evolution is indeed exact with respect to the constraints on the gauge fields: the theory is reformulated such that these constraints are internal symmetries of the discrete theory.


\section{Numerical Simulations}

Our work extends the development of the \textsc{Grid and Bubble Evolver} (\textsc{GABE})~\cite{Child:2013ria,gabe}, which has been used to study a variety of $U(1)$ preheating models~\cite{Deskins:2013lfx,Adshead:2015pva,Adshead:2016iae,Giblin:2017wlo}.
The implementation of the lattice gauge variables requires a significant reworking of the numerical procedure.
Our scheme retains the typical finite-differencing approach to the numerical integration of the continuum scalar field $\phi$, which we must ensure is compatible with the link update rules.
The update rules, though they are exact, resemble a first-order integration method such as Euler's method, which is insufficient to evolve the scalar field.
Such methods fail to produce convergent evolutions of the inflaton's second-order differential equation, which is the reason \textsc{gabe} typically implements a second-order Runge-Kutta scheme.
However, this method relies on a ``guess'' of the field configurations at the midpoint of the current and next time slices, which is not well defined for the lattice variables.
To solve this problem, we update the links and electric fields with a time step half as large as that used in the second-order Runge-Kutta scheme that evolves the inflation (i.e., $a_0/2$).
This provides a way to calculate the gauge coupling terms at the ``midpoint.''

\subsection{Initial conditions}\label{sec:inits}

We initialize all of our simulations at the end of inflation, just as the inflaton enters a period of near-coherent oscillation about the minimum of its potential, \cref{infpot}.
We choose values of the field and field derivative which correspond to a point just after inflation ends, which are $\phi_0 \approx .193 \, m_\text{pl}$ and $\phi_0' \approx -14.2 \,m_\phi {m_\text{pl}}$.
These values depend on the shape of the potential, \cref{infpot}, yet are independent of the choice of inflationary scale $m_\phi$ (although we do not vary this parameter here).
The fluctuations of the inflaton are assumed to be Bunch-Davies~\cite{Easther:2010mr,Jedamzik:2010dq},
\begin{align}
	\left\langle \phi(k) \phi(q)^\ast \right\rangle = \frac{1}{2 \omega(k)} \delta^{(3)}(k-q).
\end{align}
The homogeneous modes of the continuum gauge fields are initialized to zero, with fluctuations likewise given by
\begin{align}
	\left\langle A_i^a(k) A_j^b(q)^\ast \right\rangle &= \frac{1}{2 \omega(k) \Theta(\phi_0)} \delta^{(3)}(k-q) \delta_{ab} \delta_{ij}.
\end{align}
The frequency is defined by $\omega(k)^2 = k^2 + m_\text{eff}^2$, where for the scalar $\phi$ the effective mass $m_\text{eff} = \partial^2 V / \partial \phi^2$.
The gauge fields are massless, so $\omega(k) = \abs{k}$.
Note the appearance of the initial coupling coefficient, $\Theta(\phi_0)$, which ensures that the gauge kinetic term is canonically normalized~\cite{Caldwell:2011ra,Motta:2012rn,Maleknejad:2012fw}.

Higher frequency modes, with large gradients between lattice sites, are highly susceptible to numerical noise.
For this reason, we introduce a momentum space window function, 
\begin{align}
	F(k) = \frac{1 - \tanh\left( s(k - k_\ast) \right)}{2},
	\label{windowfun}
\end{align}
where the parameter $s$ sets the sharpness of the damping edge and $q$ sets the frequency cutoff.
While the lattice spacing sets a natural ultraviolet cutoff scale, we expect most of the resonance to occur near the infrared.
Therefore, damping the upper-UV in high resolution simulations (with large Nyquist frequencies) allows us to limit numerical error accrued from nonphysical sources while still resolving the important modes.

In principle, setting initial conditions for the lattice gauge variables requires no further action than translating the continuum initial conditions to the links and electric fields by applying the standard identity,
\begin{align}
	\exp\left[ 2 A^a \sigma^a \right] = \cos\abs{A} \sigma^0 + 2 \frac{A^a}{\abs{A}} \sin\abs{A} \sigma^a,
\end{align}
which in turn sets the flavor coefficients of 
\begin{align}
	E = \exp\left[ 2 A^a \sigma^a \right] \exp\left[ 2 B^b \sigma^b \right]
\end{align}
to
\begin{align}
	E^0 &= \cos\abs{A} \cos\abs{B} - \frac{ A^a B^a }{\abs{A} \abs{B}} \sin\abs{A} \sin\abs{B},
\end{align}
and
\begin{align}
	\begin{split}
		E^a &= 2 \Bigg[ \frac{ A^a }{\abs{A}} \sin\abs{A} \cos\abs{B} + \frac{ B^a }{\abs{B}} \cos\abs{A} \sin\abs{B} \\
		&\hphantom{={} 2 \Bigg[ }
		- \epsilon^{abc} \frac{ A^b B^c }{\abs{A} \abs{B}} \sin\abs{A} \sin\abs{B} \Bigg].
	\end{split}
\end{align}
However, careful attention must be paid to Gauss's law, which is the result of varying the action with respect to the temporal gauge fields, $A_0^a$ or $U_0$.
In Lorenz gauge, Gauss's law is a dynamical, second-order equation of motion for $A_0$ which may be integrated as any other field.
The temporal gauge instead sets both $A_0^a$ and $A_0^a{}'$ to zero, which sets $U_0$ and $E_0$ to the identity.
While the choice of temporal gauge can be trivially satisfied on the lattice, we have no guarantee that our initial conditions satisfy Gauss's law, which in this gauge is a constraint on the time derivative of the vector potential, $A_i^a{}'$, on spatial slices.

The continuum version of Gauss's law is
\begin{align}\label{eqn:ctmgauss}
	\partial^i A_i^a{}' + \frac{d \Theta/d\phi}{\Theta} \partial^i \phi A_i^a{}' + f^{abc} A^i_b A_i^c{}' = 0.
\end{align}
In the limit of small coupling, the constraint in \cref{eqn:ctmgauss} is approximately satisfied by setting $A_i^a{}'$ to be transverse, i.e., $\partial^i A_i^a{}' = 0$. Longitudinal modes may easily be removed with the projection operator
\begin{align}
	P_{i}^{\hphantom{i}j}(k) &= \left(\delta_i^{\hphantom{i}j} - \frac{k_i k^j}{\abs{k}}\right).
\end{align}
In the Abelian case, with relatively weak coupling to the scalar inflaton, applying this operator on the initial slice is a sufficient approximation.
However, for non-Abelian fields, Gauss's law is increasingly violated by stronger couplings, whether to the scalar or to the other gauge flavors.

Though the lattice evolution preserves the satisfaction (or the degree of violation) of Gauss's law exactly, configurations which violate it exhibit numerical instability in the form of unphysical growth of the electric fields~\cite{Moore:1996wn,Moore:1996qs}.
Some prior non-Abelian preheating studies have addressed Gauss's law on the initial slice with a period of ``cooling'' or dissipation~\cite{Rajantie:2000nj}, or with Monte-Carlo sampling~\cite{Ambjorn:1990wn,Ambjorn:1990pu,Smit:2002yg,Skullerud:2003ki}.
In our case, Gauss's law may also be trivially solved by setting the individual $A_i^a{}'$ to zero (which in turn sets each $E_i$ to the identity matrix), as done in Ref.s~\cite{Enqvist:2015sua,Tranberg:2017lrx}.
Alternately, setting all the $A_i^a = 0$ on the initial slices removes the non-Abelian term from Gauss's law, and in some cases allows the resulting constraint to be satisfied in momentum space~\cite{Tranberg:2003gi,GarciaBellido:2003wd}.
(Because Gauss's law for our model is implicit, such a prescription is not applicable.)
However, these choices violate the uncertainty principle, breaking down the semiclassical approximation of the Bunch-Davies vacuum.
Therefore, producing nontrivial, self-consistent initial conditions which satisfy Gauss's law is critical.
For this reason, we will employ a dissipative method similar to~\cite{Rajantie:2000nj} as detailed below (and in \cref{Arelax}) and compare the results to simulations beginning with trivial (i.e., $A_i^a{}' = 0$) initial conditions.

To set nontrivial initial conditions, we begin by projecting both the fields, $A_i^a$ (by our remaining gauge freedom), and their time derivatives, $A_i^a{}'$, to be transverse.
Next, we define the local Gauss constraints $G^a(x)$ ($a \in \{1,2,3\}$) by
\begin{align}
	G^a &\equiv \sum_{i\neq 0} \left( \Theta(x) E_i^a(x) - \Theta(x-\hat{\imath}) \tilde{E}_i^a(x) \right),
\end{align}
where $\tilde{E}$ represents the parallel-transported electric field
\begin{align}
	\tilde{E}_i^a \equiv \text{Tr}\left[ - 2 \sigma^a U_i^\dag(x-\hat{\imath}) E_i(x-\hat{\imath}) U_i(x-\hat{\imath}) \right].
\end{align}
As Gauss's law requires $G^a(x) = 0$, we implement an iterative relaxation method seeking to minimize the ``Hamiltonian''~\cite{Rajantie:2000nj,Ambjorn:1990pu,Ambjorn:1990wn}
\begin{align}
	H = \sum_x G^a(x) G^a(x).
\end{align}
This amounts to evolving the dissipative equations
\begin{align}
	\frac{\partial E_j^c(x)}{\partial t} = - \frac{\partial H}{\partial E_j^c(x)},
\end{align}
which evolve the electric fields in the direction which minimizes the global Gauss violation $H$ (see \cref{Arelax} for details on the numerical implementation).
In practice, it is impossible to relax the fields until the grid-averaged Gauss violation is zero to machine precision.
We instead relax until the Gauss violation is some fraction (smaller than $10^{-5}$) of the originally initialized value, testing multiple such fractions to ensure that further relaxation has no effect on the resulting evolution.

\subsection{Lattice parameters}

All simulations presented use lattices with $256^3$ points and an initial box length $L= 5 \, m_\phi^{-1} \approx 2.5 \, \mathcal{H}_0^{-1}$, where $\mathcal{H}_0$ is the Hubble parameter at the beginning of the simulation, and the lattice spacing ratio is $\kappa = a_0 / a_i = 1/20$.
As mentioned above, the degree to which our simulations satisfy Gauss's law is constant to numerical precision, supporting the robustness of our implementation~\cite{Ambjorn:1990pu,Rajantie:2000nj}.
The Nyquist frequency of this lattice is
\begin{align}
	k_\text{nq} \equiv 256 \sqrt{3} \frac{2 \pi}{L}
	\approx 557 \, m_\phi,
\end{align}
and we apply the momentum-space window function, \cref{windowfun}, to our initial conditions with cutoff $k_\ast = k_\text{nq}/4$ and smoothness parameter
\begin{align}
	s = \frac{1}{4} \frac{L}{2\pi}
	\approx .2 \, m_\phi^{-1}.
\end{align}
Such a window function was imposed in~\cite{Deskins:2013lfx} to reduce the effect of numerical noise on discrete spatial derivatives as a means to improve the satisfaction of gauge conditions. 
The lattice formulation (working in the temporal gauge) does not precipitate such a need; however, we retain the use of the window function because, as discussed above, with our gauge choice Gauss's law is now a differential constraint on spatial slices of the gauge fields.
Imposing the window function reduces the degree of violation of Gauss's law incurred by the initial conditions when generated in momentum space, which in turn reduces the computational time required by the relaxation method.

As discussed below, the lattice variables require particularly high spatial resolution.
For this reason, our chosen box length is relatively small compared to other studies.
Another consequence of shrinking the box length is the addition of higher frequency modes.
As the size of the box decreases, the contribution of fluctuations to the energy density increases relative to the homogeneous component, since energy in the gauge modes scales with the cutoff scale to the fourth power, $k_\ast^4$. 
This energy, which arises as a result of treating the vacuum as a classical field configuration, is not physical. 
The contribution of these zero-point fluctuations to the energy density is typically ignored in lattice treatments, provided their contribution to the total energy remains small. 
Imposing a cutoff ensures that this is the case without eliminating the necessary vacuum ``seed" at the relevant scales.
We note that the resonance band is always well within the initial cutoff $k_\ast$, meaning we suppress no modes which are relevant during resonance.

If the physically relevant modes lie far from the Nyquist frequency of our box, one might question the need for utilizing grid sizes upwards of $256^3$.
Simulations of scalar fields (and continuum gauge fields) require sufficient spatial and temporal resolution to accurately resolve any dynamic modes, which amounts to at least ten time slices per oscillation of the highest-frequency mode.
Issues of numerical resolution are even more critical in the case of continuum simulations of gauge fields, where the satisfaction of gauge constraints (which are often differential constraints) is not guaranteed to be preserved by numerical evolution. These cases demand as much precision as computationally feasible.

In contrast, the lattice formulation provides an evolution which exactly preserves the gauge constraints, regardless of the lattice dimensions.
As discussed above, the role of numerical resolution appears instead at the level of the theory, rather than the integration: the lattice variables compactify the gauge group.
That is, the link variables are (matrix-valued) rotations with phase argument proportional to products of the lattice spacings, the gauge internal coupling, and the vector potentials, i.e., $a_i g A_i^a$.
Likewise, the electric fields depend upon the combination $a_0 a_i g A_i^a{}'$.
These phase variables lie within the compact domain $(-\pi,\pi]$, while the gauge potentials themselves are unbounded.
Therefore, in our simulations high numerical resolution is not only required to accurately evolve the scalar, but also to recover the full domain taken on by the gauge fields.
Since during preheating the gauge fields are strongly amplified, this requirement is significant.
Our choice of box length, $L = 5 \, m_\phi^{-1}$, strikes a compromise between the resolution requirements and the inclusion of sufficient IR modes and is still larger than the horizon at the end of resonance.


\section{Results}

To set our expectations for the nonperturbative decay of the inflaton, we first look for the approximate location of potential tachyonic instabilities.
At the end of inflation, the inflaton's dynamics are dominated by the oscillation of its homogeneous mode about the minimum of its potential, and so (for a short period of time) takes the form
\begin{align}
	\phi(\tau) = \phi_0 \cos( m_\phi \tau).
\end{align}
In the case where the non-Abelian contributions are small, the mode amplitude $A_i(k, \tau)$ obeys the linearized equation of motion
\begin{align}\label{linearEOM}
	A_i''(k, \tau) + k^2 A_i(k, \tau) &\approx \frac{\phi_0 m_\phi}{M} \sin(m_\phi \tau) F_{0i}(k, \tau)
\end{align}
in the Lorenz gauge.
Initially (regardless of gauge choice) we set $A_0 = 0$, so that $F_{0i} \approx A_i'$ early in the simulation.
Tachyonic resonance will occur if there are values of $k$ for which the mode equation, \cref{linearEOM}, has an imaginary frequency.
Assuming that the sine function is maximally positive, we can seek exponential amplification via the WKB approximation,
\begin{align}
	A_i(k, \tau) \sim A_i(k) \exp\left( i \int^\tau \omega_k(\tau') d\tau' \right),
\end{align}
which holds when $\dot{\omega}_k \gg \omega_k^2$.
In this limit, \cref{linearEOM} takes the form
\begin{align}\label{WKBeom}
	\left( - \omega_k(\tau)^2 + k^2 - \frac{ \phi_0'(\tau)}{M} i \omega_k(\tau) \right) A_i(k) = 0.
\end{align}
The characteristic equation corresponding to \cref{WKBeom} exhibits that solutions $A_i(k, \tau)$ with a negative imaginary component of $\omega_k$ (i.e., exponentially growing solutions) are strongest within the band
\begin{align}
	k^2 < \frac{\phi_0^2 m_\phi^2 }{ 4 M^2 } \approx \left( 6 \, m_\phi \right)^2,
\end{align}
where the final relation holds for our typical parameters $M = .016 \, m_\text{pl}$, $\phi_0 = .193 \, m_\text{pl}$, and $m_\phi = 10^{-6} \, m_\text{pl}$.

\subsection{The Abelian limit}

\begin{figure}[htbp]
\centering
\includegraphics[width=.99\columnwidth]{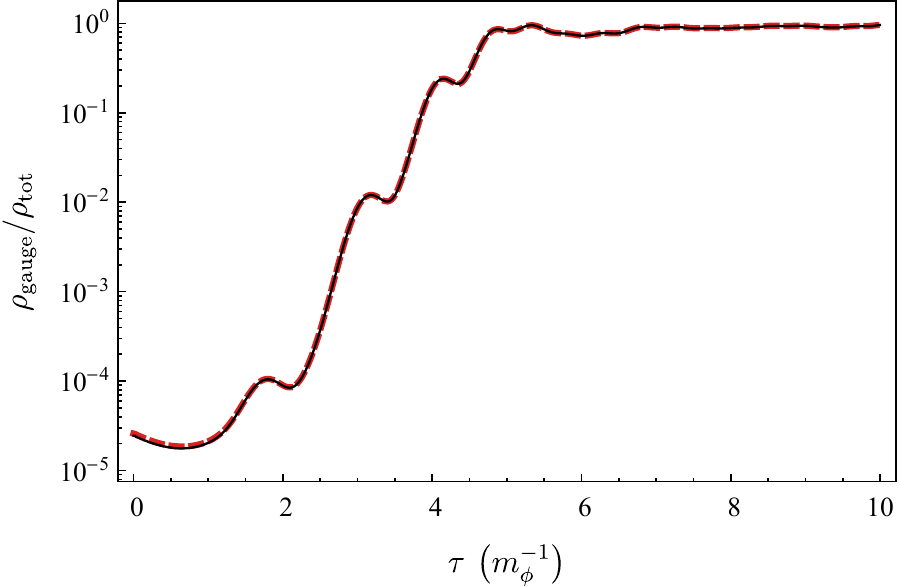} \\
\vspace{1eX}
\includegraphics[width=.99\columnwidth]{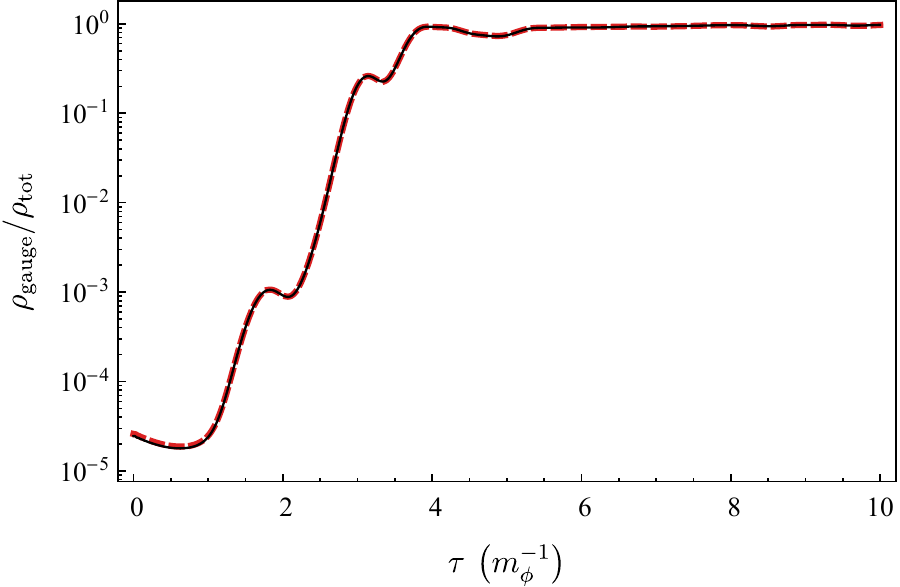}
\caption{
	The evolution of the energy fraction $\rho_\text{gauge}/\rho_\text{tot}$ for two coupling strengths: $M = .016 \, m_\text{pl}$ (top) and $M = .014 \, m_\text{pl}$ (bottom).
	In each panel the solid black curve tracks this ratio for a continuum simulation of three $U(1)$ gauge fields and the dashed red curve corresponds to a simulation of $SU(2)$ lattice fields, with $g = 10^{-6}$, using the procedure described in the text.
}\label{fig:nearAbelian}
\end{figure}

We begin by corroborating the results of our lattice-gauge simulations with the $U(1)$ continuum simulations that appeared in~\cite{Deskins:2013lfx}.
In \cref{fig:nearAbelian}, we show results of the simulation of three continuum $U(1)$ fields each with identical dilatonic couplings and compare to a simulation using the lattice $SU(2)$ variables, where the three $SU(2)$ flavors are very weakly coupled to each other ($g = 10^{-6}$).
The comparison demonstrates both that our lattice update scheme correctly reproduces the continuum evolution and that our relaxation method realizes initial conditions which remove any unphysical instabilities from violations of Gauss's law (see \cref{fig:moreRelax} of \cref{Arelax} for a depiction of such instability present in simulations with less applied relaxation).
\Cref{fig:nearAbelian} depicts that the evolutions of the energy fraction $\rho_\text{gauge} / \rho_\text{tot}$ from both simulations are nearly identical for two values of $M$.
For $M = .016 \, m_\text{pl}$, $96.2\%$ of the final energy density resides within the gauge fields implementing the lattice variables, compared to $95.7\%$ for the continuum software; for $M = .014 \, m_\text{pl}$ these figures are $97.5\%$ and $97.4\%$, respectively.

\subsection{Non-Abelian preheating}

\begin{figure}[thbp]
\centering
\includegraphics[width=.99\columnwidth]{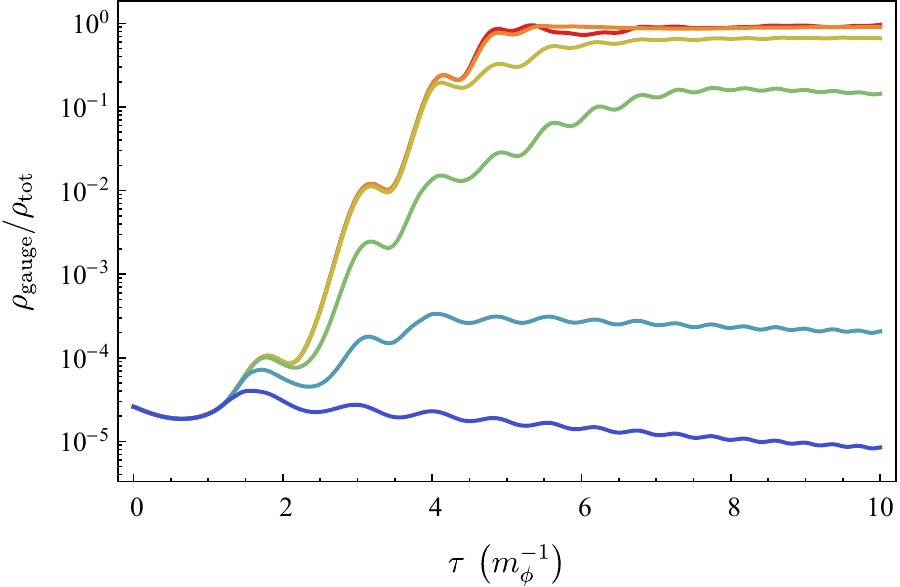} \\
\vspace{1eX}
\includegraphics[width=.99\columnwidth]{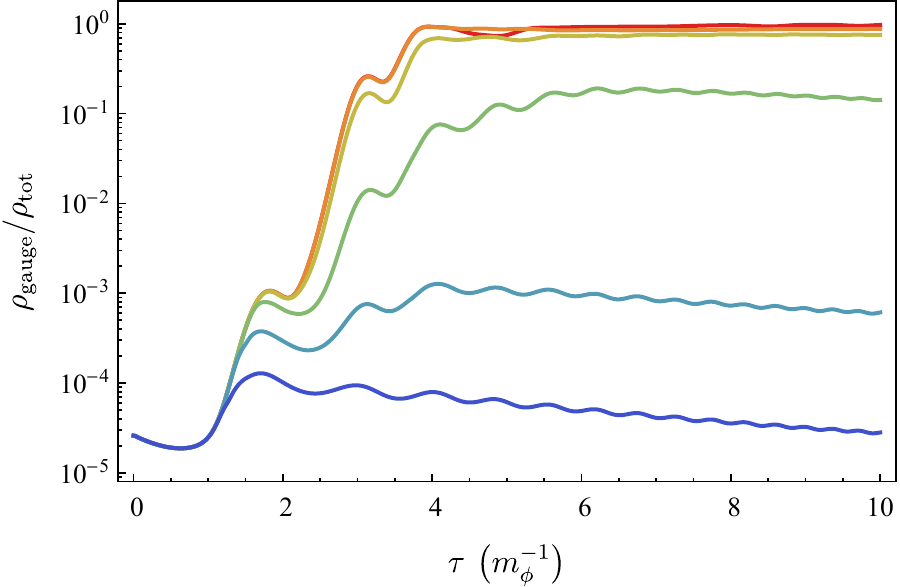}
\caption{
	The evolution of the energy fraction $\rho_\text{gauge}/\rho_\text{tot}$ for inflaton couplings $M = .016 \, m_\text{pl}$ (top) and $M = .014 \, m_\text{pl}$ (bottom), each with non-Abelian coupling strengths $g = 10^{-6}$ (red, topmost line in each panel) to $g = 10^{-1}$ (blue, bottommost line in each panel), by factors of $10$.
}\label{fig:energyFracs}
\end{figure}

Having established the efficacy of our software for weakly coupled gauge fields, we now turn to the effect of the non-Abelian coupling strength on preheating under various strengths of coupling to the inflaton.
In \cref{fig:energyFracs} we show the effect of increasing the self-coupling, $g$, for two different values of the coupling to the inflaton sector, $M$.

\begin{figure}[htbp]
\centering
\includegraphics[width=.99\columnwidth]{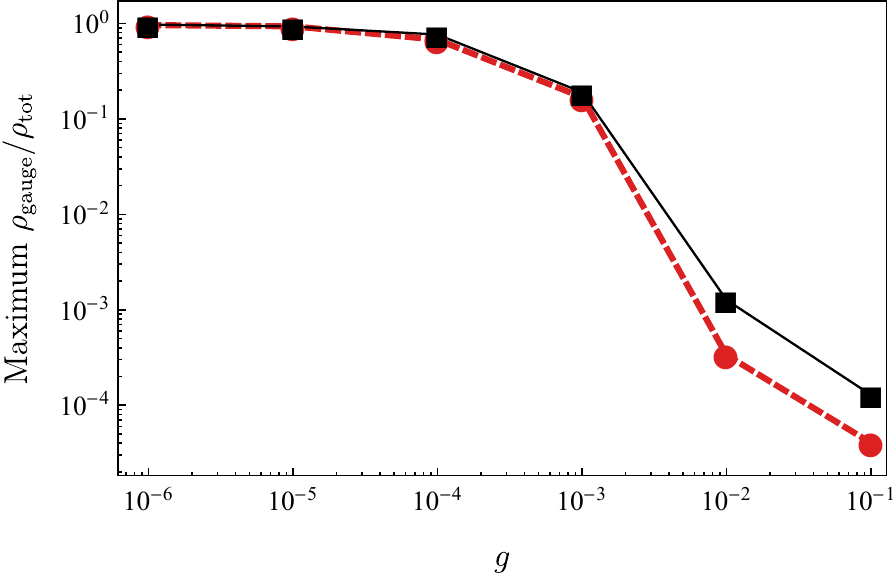}
\caption{
	The maximum energy fraction $\rho_\text{gauge}/\rho_\text{tot}$ vs. $g$ for $M = .016 \, m_\text{pl}$ (dashed red, circles) and $M = .014 \, m_\text{pl}$ (solid black, squares) with nontrivial initial conditions.
}\label{fig:moneyPlots}
\end{figure}

It is immediately evident in \cref{fig:energyFracs} that increasing the strength of non-Abelian coupling drastically diminishes the efficiency of preheating.
Indeed, at near-Standard Model values ($g \sim .1$), less than $.001\%$ of the energy in the lattice resides in the gauge sector at the end of the simulations ($\tau = 10 \, m_\phi^{-1}$).
In \cref{fig:moneyPlots} we examine the success of preheating as a function of the self-coupling $g$ by plotting the maximal value of $\rho_\text{gauge}/\rho_\text{tot}$ for the simulations depicted in \cref{fig:energyFracs}.

To understand this parametric suppression, return to \cref{linearEOM}: one way to incorporate the effects of the non-Abelian interactions is by considering terms which resemble $g^2 A_j A^j A_i$ as a mass term $\sim m^2 A_i$, which naturally shifts the upper limit of resonance via
\begin{align}
	k^2 + m^2 < \frac{\phi_0^2 m_\phi^2 }{ 4 M^2 }.
\end{align}
The effect of this time dependent mass becomes significant once $g A_i$ is of order one, which is clearly reached earlier as $g$ increases.
This reasoning aligns with the results of \cref{fig:energyFracs}, where $\rho_\text{gauge}$ follows a similar trajectory early on until preheating shuts off, which occurs sooner as $g$ increases.

\begin{figure}[thbp]
\centering
\includegraphics[width=.99\columnwidth]{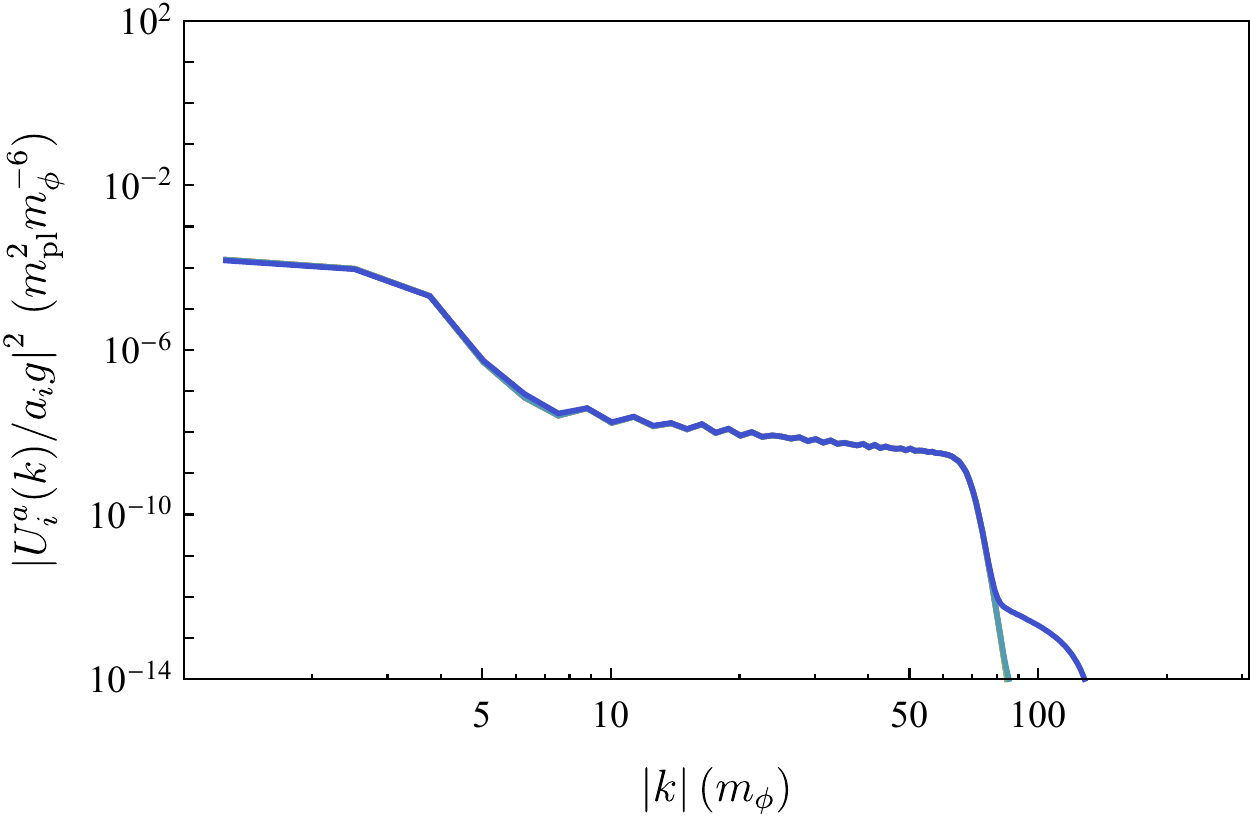} \\
\vspace{1eX}
\includegraphics[width=.99\columnwidth]{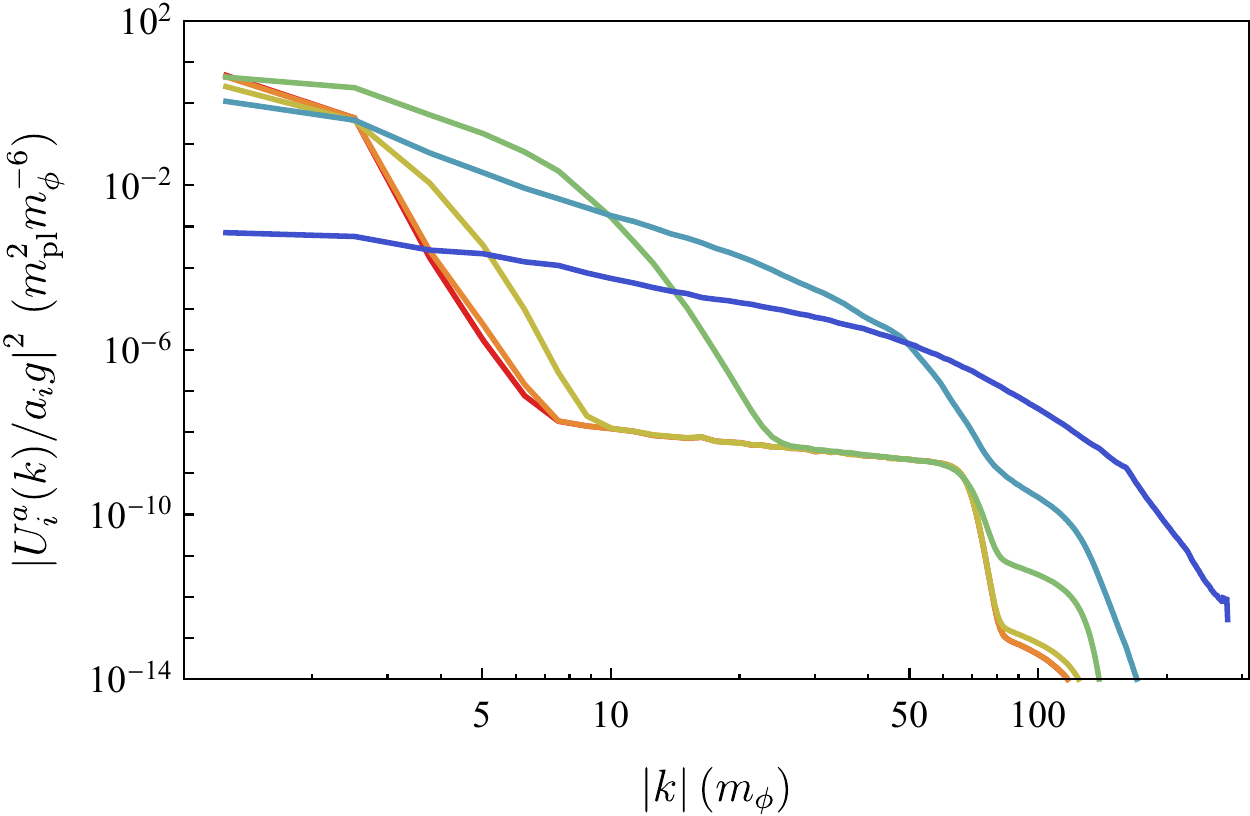}
\caption{
	The power spectra of an arbitrary link field $U_i^a(k)$, scaled by $a_i g$ for $M = .016 \, m_\text{pl}$.
	The top panel shows the effect of the initial resonance near the first zero crossing of the inflaton ($\tau = 1.25 \, m_\phi^{-1}$) which is near identical among all values of $g$.
	The bottom panel, depicting the spectra near the third zero crossing ($\tau = 3.75 \, m_\phi^{-1}$), demonstrates that the non-Abelian effects (rescattering) occur earlier for larger values of $g$.
	The colors in both panelsc correspond to $g = 10^{-6}$ (red, bottommost when applicable) through $g = 10^{-1}$ (blue, topmost when applicable), spaced by factors of $10$.}
\label{fig:fieldSpecEvolution}
\end{figure}

\begin{figure}[thbp]
\centering
\includegraphics[width=.99\columnwidth]{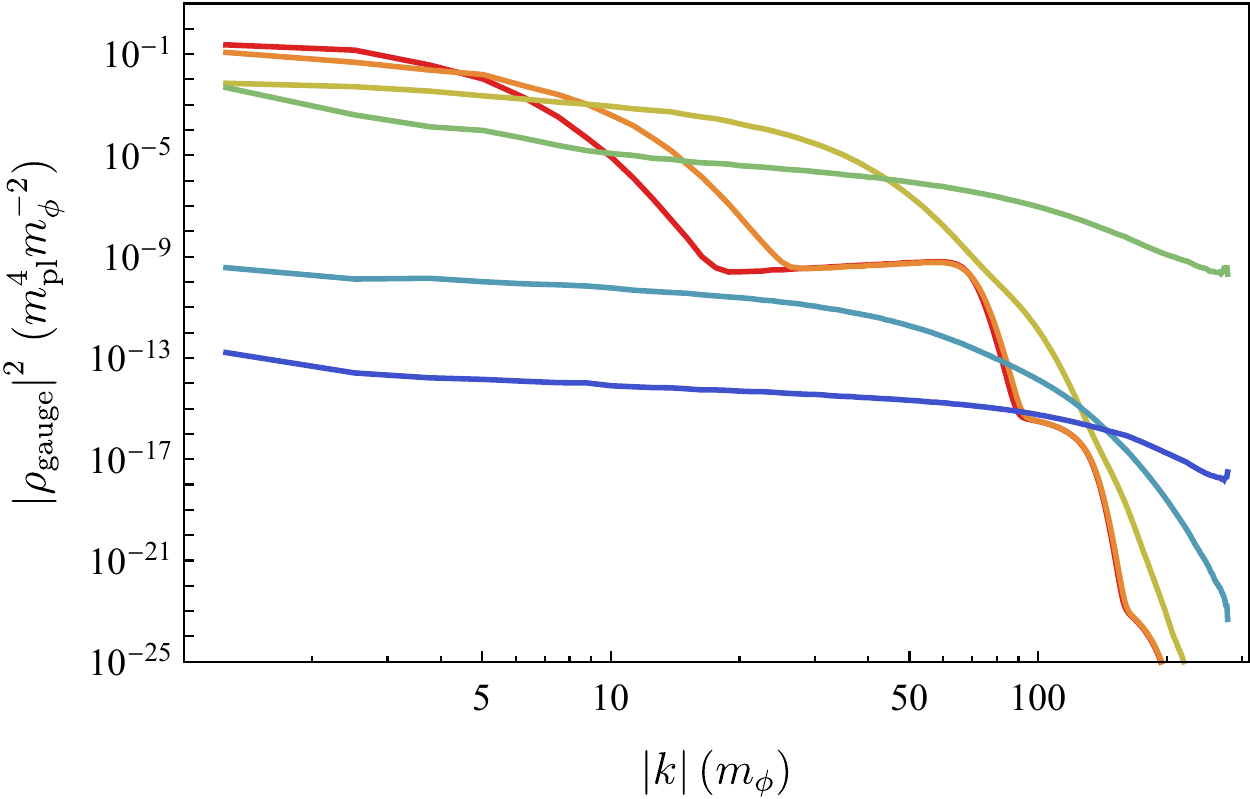} \\
\vspace{1eX}
\includegraphics[width=.99\columnwidth]{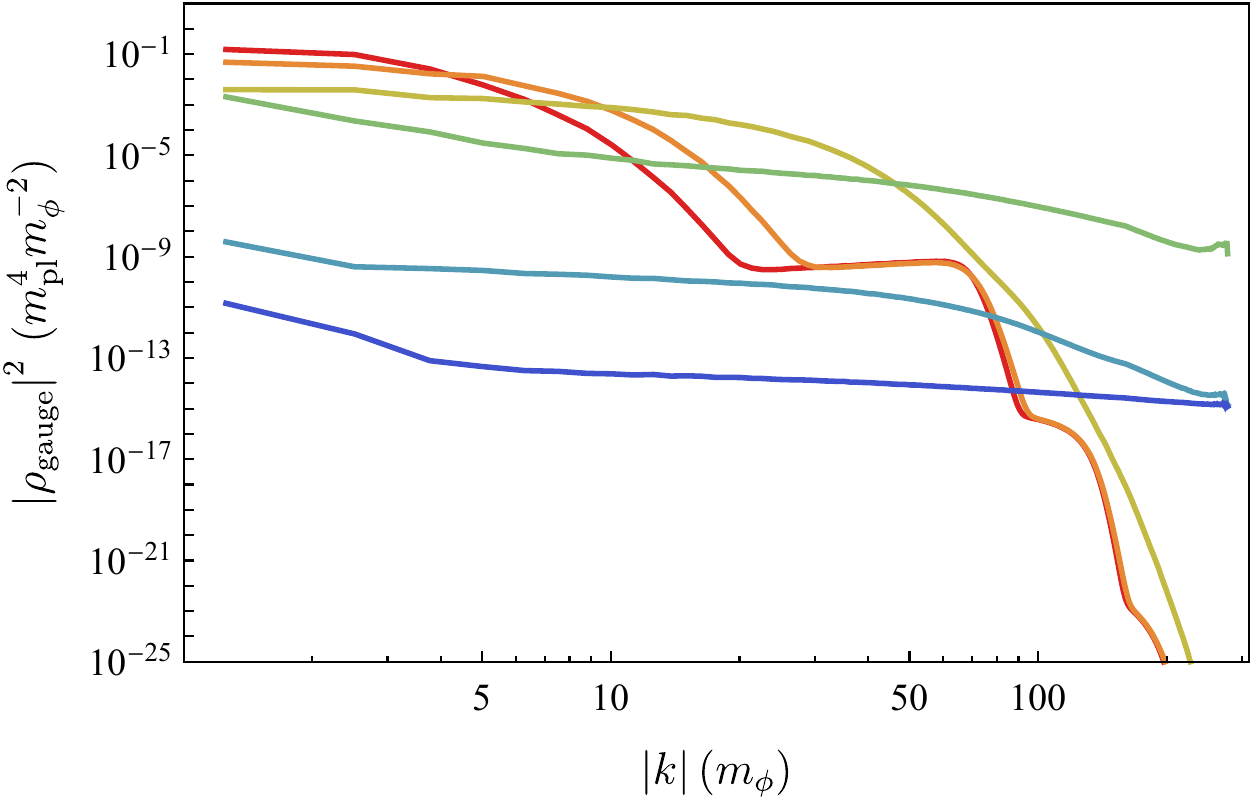}
\caption{
	Power spectra of the magnetic field energy density $\rho_{B}(k)$ at the end of the simulations ($\tau = 10 \, m_\phi^{-1}$) for inflaton couplings $M = .016 \, m_\text{pl}$ (top) and $M = .014 \, m_\text{pl}$ (bottom), each with non-Abelian coupling strengths $g = 10^{-6}$ to $g = 10^{-1}$ (red through blue), spaced by factors of $10$.
}\label{fig:finalSpectra}
\end{figure}

We can also track the suppression of preheating by examining the field spectra, as in \cref{fig:fieldSpecEvolution}.
The effect of the non-Abelian interactions is not limited to blocking resonance: once the non-Abelian interactions dominate the dynamics of the gauge fields, the energy density is spread to larger wavenumbers.
\Cref{fig:finalSpectra} demonstrates that increasing $g$ pushes power further and further out into the ultraviolet.
This cascade is another reason our simulations require high resolution.
While resonance always occurs well within the cutoff $k_\ast$, the gauge bosons subsequently rescatter into higher-$k$ modes; these nonlinear dynamics must also be well resolved.

\subsection{Trivial Gauss configurations}

\begin{figure}[htbp]
\centering
\includegraphics[width=.99\columnwidth]{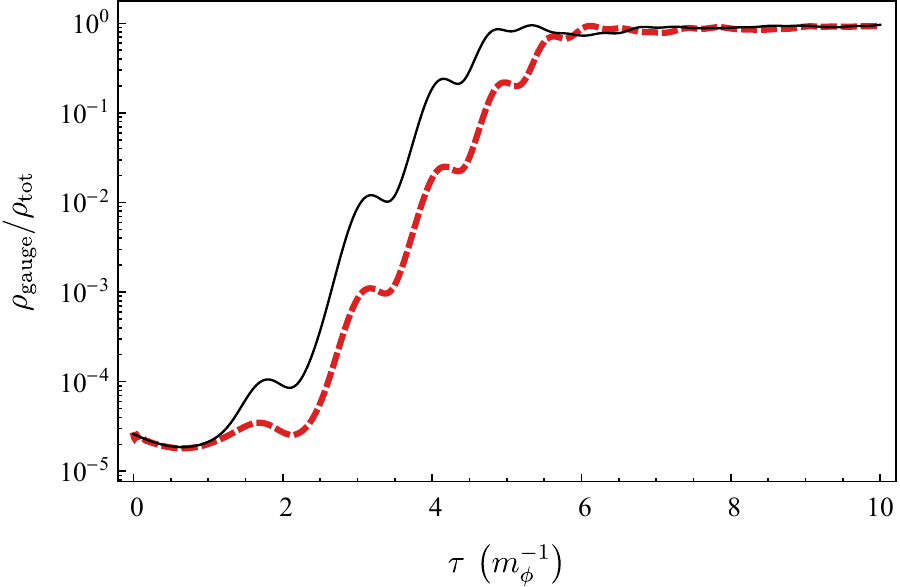} \\
\vspace{1eX}
\includegraphics[width=.99\columnwidth]{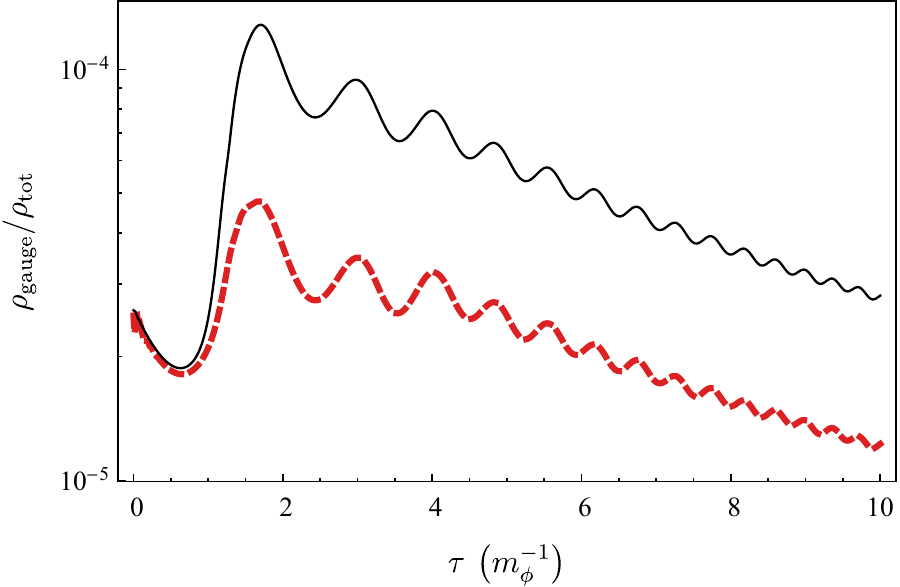}
\caption{
	The evolution of the energy fraction $\rho_\text{gauge}/\rho_\text{tot}$ for $M = .016 \, m_\text{pl}$ and $g = 10^{-6}$ (top) and $M = .014 \, m_\text{pl}$ and $g = 10^{-1}$ (bottom) with trivial (red, dashed) and nontrivial (solid black) initial configurations of the gauge fields.
}\label{fig:trivialCompare}
\end{figure}

In this section, we quantify the effect of initializing nontrivial initial conditions (namely, those which satisfy Gauss's law via relaxation) compared to trivial ones.
By setting the initial electric fields to zero [namely, $A_i'(k) = 0$] and doubling the initial magnetic field strength, we evolve a configuration which satisfies Gauss's law to machine precision via a trivial field configuration while preserving the total initialized energy in the gauge sector.
\Cref{fig:trivialCompare} compares such a simulation to the evolution of a nontrivial configuration for various values of $g$ and $M$.

\begin{figure}[htbp]
\centering
\includegraphics[width=.99\columnwidth]{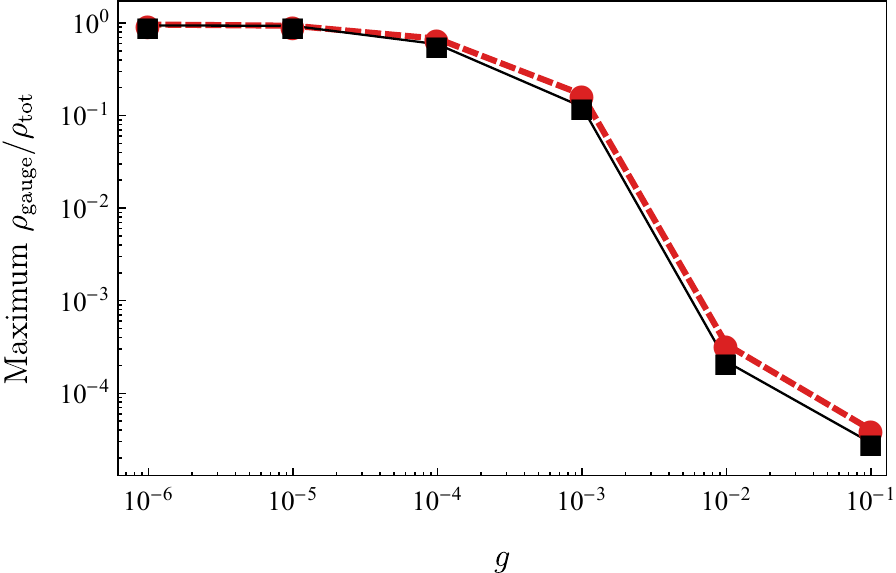}
\caption{
	The maximum energy fraction $\rho_\text{gauge}/\rho_\text{tot}$ vs. $g$ for $M = .016 \, m_\text{pl}$ with trivial (dashed red, circles) and nontrivial (solid black, squares) initial conditions.
}\label{fig:trivMoneyPlots}
\end{figure}

\Cref{fig:trivMoneyPlots} depicts the success of preheating for trivially initialized electric fields relative to nontrivial initial conditions, again via the maximum achieved preheating fraction $\rho_\text{gauge} / \rho_\text{tot}$.
These results verify that our results are only insensitive to the precise details of the initial configuration in the region of parameter space where preheating is successful.
The discrepancy becomes significant for larger $g$, though this region of parameter space is uninteresting with respect to preheating.
However, our results suggest that, for non-Abelian models with tachyonic instabilities, the difference between trivial and nontrivial initial conditions may be substantial.

\subsection{Negative initial field values}

\begin{figure}[htbp]
\centering
\includegraphics[width=.99\columnwidth]{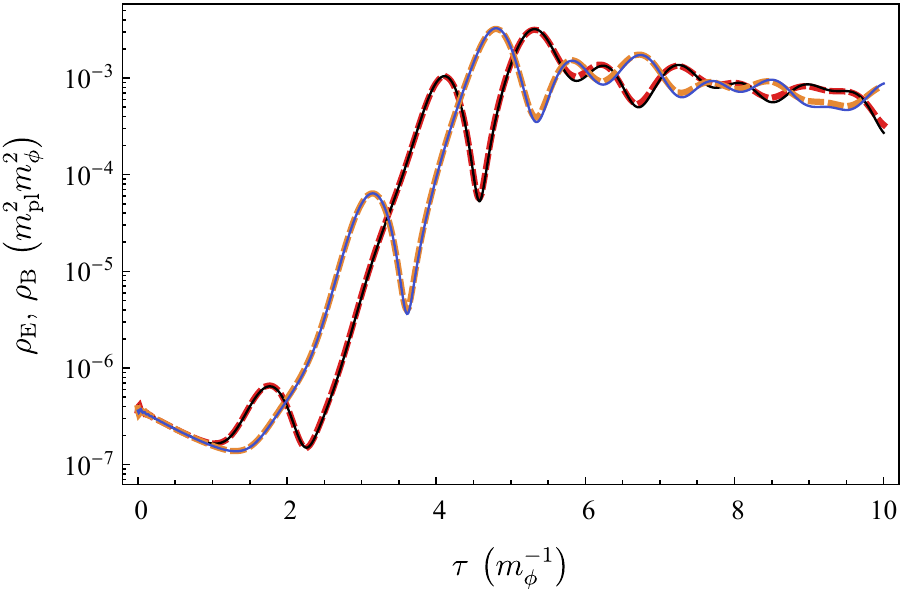} \\
\vspace{1eX}
\includegraphics[width=.99\columnwidth]{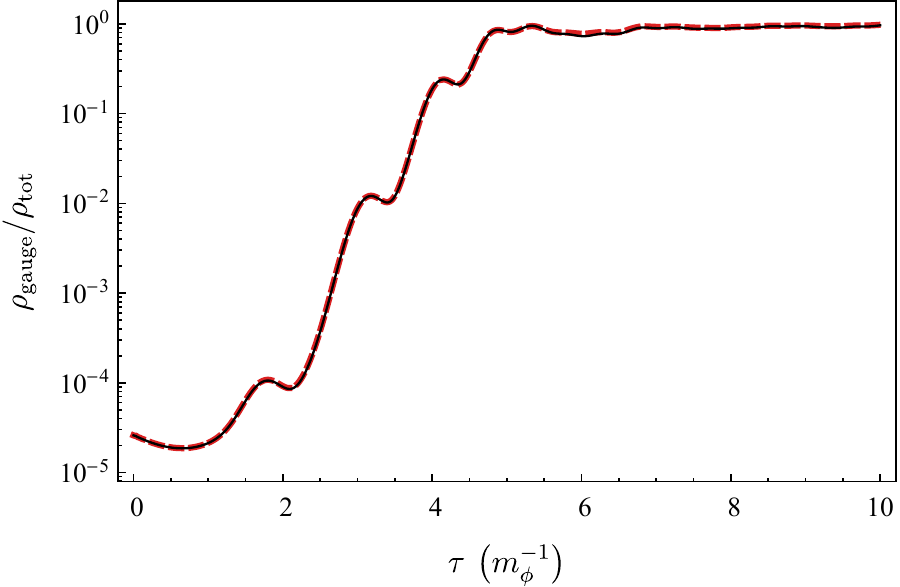} \\
\vspace{1eX}
\includegraphics[width=.99\columnwidth]{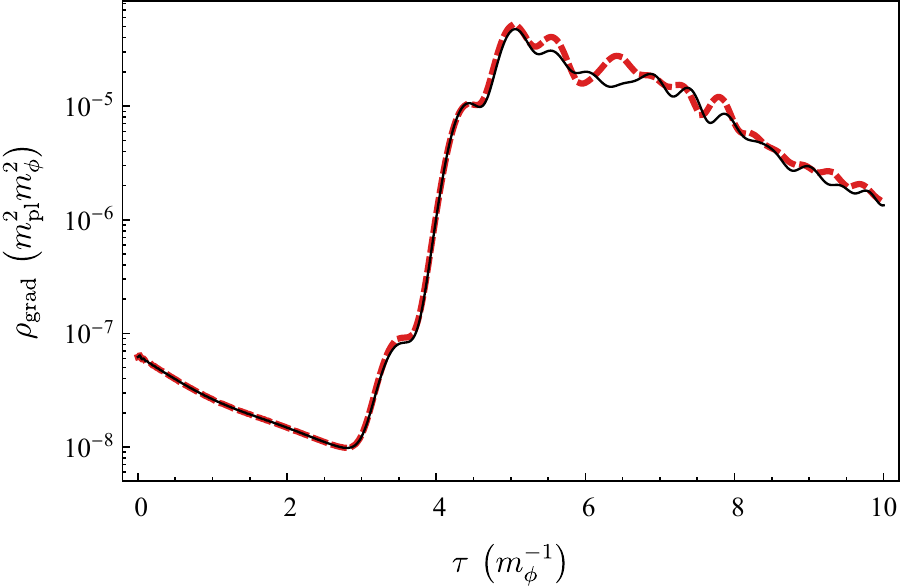}
\caption{
	A comparison of evolutions for positive and negative initial values of $\phi_0$ for couplings $M = .016 \, m_\text{pl}$ and $g = 10^{-6}$.
	The top panel shows the difference between $\phi_0 > 0$, where $\rho_E$ (solid black) follows a similar trajectory to the negative-coupling $\rho_B$ (dashed red), while the positive-coupling $\rho_B$ (solid blue) aligns with the negative-coupling $\rho_E$ (dashed orange).
	The middle panel demonstrates that the evolution of the overall gauge energy fraction $\rho_\text{gauge} / \rho_\text{tot}$ is largely unchanged when interchanging $\phi_0 > 0$ (solid black) for $\phi_0 < 0$ (dashed red).
	The most substantial (though still marginal) difference lies in the backreaction onto the scalar gradient energy $\rho_\text{grad}$ in the bottom panel, again for $\phi_0 > 0$ (solid black) and $\phi_0 < 0$ (dashed red).
}\label{fig:rhoEBswitch}
\end{figure}

We now present the results for simulations where the field comes in from $\phi< 0$ [i.e., where the initial strength of the coupling is $\exp(-\abs{\phi_0}/M)$].
\Cref{fig:rhoEBswitch} demonstrates that the most significant effect is that the magnetic fields are the first to be enhanced, rather than the electric fields in the positive coupling case.
In fact, the evolutions of $\rho_E$ for positive $M$ and $\rho_B$ for negative $M$ (and vice versa) match almost exactly through the first oscillation or so.
The trajectory of the net gauge field energy density is largely unchanged, though the subsequent backreaction onto the scalar is slightly modified.

We find that the maximal fraction $\rho_\text{gauge}/\rho_\text{tot}$ changes by $\lesssim 4\%$ (often subpercent) in all cases that were numerically stable.
For negative values of $M$, our simulations are even more constrained by the limits of the discretization because the initial gauge field amplitudes are scaled by $1/\sqrt{\Theta(\phi_0)}$.
This normalization changes by some five orders of magnitude when changing the sign of $M$, such that larger values of $g$ are inaccessible to our particular lattice configurations.
However, we take the results of runs using smaller values of $g$ to suggest that the results are insensitive to the sign of the coupling.
While it is known that the family of models ours belongs to cannot be responsible for inflationary magnetogenesis without (at some point) spoiling the inflationary background, we see here evidence for a scenario in which strongly coupled gauge fields may (1) preserve inflationary dynamics while (2) generating primordial magnetic fields during preheating.


\section{Discussion}

In this work, we have produced the first finite-time lattice simulations of $SU(2)$ gauge fields coupled to an uncharged, real scalar field in an expanding background.
This work broadens the scope of preheating studies on the lattice to non-Abelian gauge groups, a challenging but crucial step towards simulating preheating models with minimal extensions to the Standard Model.

As observed in $U(1)$ simulations~\cite{Deskins:2013lfx}, a dilatonic-type coupling offers an extremely efficient preheating channel to $SU(2)$ gauge fields.
However, tuning the non-Abelian coupling strength toward Standard Model values rapidly blocks the resonance band in which the gauge fields become populated.
This result is significant, as it suggests that tachyonic instabilities akin to that sourced by the dilatonic coupling may struggle to efficiently preheat into non-Abelian gauge fields.
In the case presented here, significantly stronger couplings to the inflaton are required to achieve a sufficiently broad band of resonance.

The software developed for this paper offers the ability to study other preheating models with $SU(2)$ fields, as well as other cosmological contexts which involve non-Abelian fields.
While we have here considered the simplest model of inflation as a massive, real scalar singlet, the ability to evolve $SU(2)$ fields in an expanding universe brings our code closer to studies of more complicated models of inflation with gauge fields~\cite{Adshead:2012kp}.


\begin{acknowledgments}

We thank Mustafa Amin, Kaloian Lozanov, and Scott Watson for useful conversations.
The work of P. A. was supported in part by a NASA Astrophysics Theory Grant No. NNX17AG48G.
J. T. G. is supported by the National Science Foundation, Grant No. PHY-1414479.
Z. J. W. is supported in part by the United States Department of Energy Computational Science Graduate Fellowship, provided under Grant No. DE-FG02-97ER25308.
This work was performed in part at Aspen Center for Physics, which is supported by National Science Foundation Grant No. PHY-1607611.
We acknowledge NASA, the National Science Foundation, and the Kenyon College Department of Physics for providing the hardware used to carry out these simulations. 

\end{acknowledgments}


\appendix


\section{Lattice gauge theory}\label{linksEOMappendix}

In this appendix we detail the application of real-time lattice gauge methods to our preheating study.
Generally, simulating field theories on a discrete lattice requires careful attention to ensure the physical results are independent of the details of the lattice itself.
For instance, the lattice must be both large enough to encompass all physically relevant modes (the ``infrared") and also fine enough to accurately resolve the shortest wavelength modes (the ``ultraviolet").
Verifying that a lattice is both large enough and of sufficient resolution requires testing evolutions on grids which have larger sidelength ($L$) or have smaller lattice spacing ($a_i$).

Gauge fields complicate lattice simulations further because their evolutions are constrained. 
While each component of the vector potential $A_\mu$ has a nontrivial Euler-Lagrange equation, one is a constraint ($A_0$). Furthermore, one must either fix a gauge choice to reduce the redundant space of field configurations to a particular, fixed gauge orbit, or alternatively work with gauge-invariant variables. 
(As an exception, the Lorenz gauge casts the Euler-Lagrange equation for $A_0$ as a dynamical equation of motion).
Gauge conditions are typically (though not always) differential constraints, which are particularly difficult to both accurately evaluate and satisfy throughout the evolution on the lattice.
Spatial derivatives on a lattice are calculated by taking finite differences between neighboring sites and carry truncation error proportional to powers of the lattice spacing $a_i$.
For this reason, there is no guarantee that a discretized evolution of gauge fields will satisfy the gauge condition, though it is possible for the satisfaction of the gauge condition to be stable through numerical evolution, as has been shown in the Abelian case.

The alternative to a discretized evolution of an exact, continuum theory is to produce a discretized theory which (1) reproduces the continuum theory in the continuum limit ($a_\mu \to 0$) and (2) may be evolved exactly.
This is precisely the aim of lattice gauge theory, which recasts the dynamical degrees of freedom such that the gauge constraints (both the gauge condition and Gauss's law) are precise internal symmetries of the theory.
As such, lattice gauge theory is the natural tool for numerical studies of gauge theories.
In the remainder of this appendix we detail the formulation of lattice gauge theory and its application to our model.
For further introduction to the topic, see textbooks such as~\cite{Creutz:1984mg} and~\cite{Rothe:1992nt}.
For alternative treatments of lattice-gauge techniques in preheating scenarios, see~\cite{Rajantie:2000nj,GarciaBellido:2003wd,DiazGil:2008tf}.

Before proceeding to detail the application of lattice gauge theory to our preheating study, we first outline our conventions.
The variables $U_\mu(x)$ lie on the ``link'' between the lattice point at $x$ and the adjacent lattice point in the $\hat{\mu}$-direction, denoted by $x+\hat{\mu}$.
As the links are members of the Lie group $SU(2)$, we will work with the decomposition onto the basis $\sigma^\mu = (\mathds{1}, \sigma^a)$ where the $\sigma^a$ are the usual Pauli matrices,
\begin{align}\label{paulidef}
	\sigma^a &= \left\{
	\frac{i}{2} \left(
	\begin{array}{cc}
	0 & 1 \\
	1 & 0
	\end{array}
	\right),
	\frac{i}{2} \left(
	\begin{array}{cc}
	0 & -i \\
	i & 0
	\end{array}
	\right),
	\frac{i}{2} \left(
	\begin{array}{cc}
	1 & 0 \\
	0 & -1
	\end{array}
	\right)
	\right\}.
\end{align}
The links are therefore defined in terms of the continuum potentials to be
\begin{align}\label{linkDefinition}
	U_\mu(x) = \exp\left( - a_\mu g A_\mu^a(x) \sigma^a \right),
\end{align}
where the sum over repeated flavor index $a$ running from 1 to 3 is implied (regardless of placement).
Additionally, we drop the Einstein summation convention when using the link variables.
The flavor coefficients are thus given by
\begin{align}
	U_\mu^a(x) &= \text{Tr}\left[- 2 \sigma^a U_\mu(x) \right]
\end{align}
and
\begin{align}
	U_\mu^0 = \frac{1}{2} \text{Tr}\left[ U_\mu(x) \right].
\end{align}
These four coefficients are the dynamical degrees of freedom of our system, which are further constrained by the unitarity of the links, $U^\dag_\mu U_\mu = 1$, or
\begin{align}\label{unitarity}
	U_\mu^0(x)^2 + \frac{1}{4} U_\mu^a(x)^2 &= 1.
\end{align}

Recall the full action for our model,
\begin{align}\label{fullAction}
	S &= \int d^4 x \sqrt{-g} \left( \frac{m_\text{pl}^2}{16\pi} R + \mathcal{L}_\phi + \mathcal{L}_G \right),
\end{align}
where the inflaton term is
\begin{align}
	\mathcal{L}_\phi = - \frac{1}{2} \partial_\mu \phi \partial^\mu \phi - \frac{1}{2} m^2 \phi^2.
\end{align}
In the continuum, the Lagrangian density describing the gauge fields is
\begin{align}
	\mathcal{L}_G = - \frac{ \Theta(\phi) }{4} F_{\mu\nu}^a F^{\mu\nu}_a.
\end{align}
We first need to establish the correspondence between the continuum and lattice theories---namely, the lattice formulation of the gauge field strength which reduces to the continuum form in the $a_\mu \to 0$ limit.
Following the steps outlined in \cref{plaqToFAppendix}, one can demonstrate that
\begin{align}\label{plaqToF}
	\text{Tr}\left[ \left( F_{\mu\nu}^a(x) \right)^2 \right] = \frac{4}{a_\mu^2 a_\nu^2 g^2} \text{Tr}\left[ 1 - P_{\mu\nu}(x) \right],
\end{align}
where again $\mu$ and $\nu$ take on fixed values, rather than following a summation convention.
\Cref{plaqToF} lends a geometric interpretation to the lattice theory: the plaquette quantifies the local curvature associated with the gauge fields.
That is, the trace $\text{Tr}\left[ 1 - P_{\mu\nu}(x) \right]$ measures the plaquette's deviation from the identity matrix, i.e., zero curvature.
Using \cref{plaqToF}, the lattice form of $\mathcal{L}_G$ is
\begin{align}\label{plaqAction}
	\begin{split}
	\mathcal{L}_{G} &= \frac{2 \Theta(x)}{a_i^4 g^2} \frac{1}{a(t)^4} \Bigg( \frac{1}{\kappa^2} \sum_i \text{Tr}\left[1 - P_{0i} \right] \\
	&\hphantom{={} \frac{2 \Theta(x)}{a_i^4 g^2} \frac{1}{a(t)^4} \Bigg(} - \sum_{j > i} \text{Tr}\left[1 - P_{ij} \right] \Bigg).
	\end{split} \\
	&= \frac{\Theta(x)}{a(t)^4} \sum_{\nu>\mu} \beta_{\mu\nu} \text{Tr}\left[1 - P_{\mu\nu} \right],
\end{align}
with $\kappa$ the lattice spacing ratio $a_0/a_i$.
The coefficients $\beta_{\gamma\mu}$ encode the prefactors of the terms in \cref{plaqAction}, i.e.,
\begin{align}
	\beta_{0i} = \frac{2}{\kappa^2 a_i^4 g^2} \quad \text{and} \quad
	\beta_{ij} = - \frac{2}{a_i^4 g^2}.
\end{align}

Our task in this Appendix is to formulate a scheme to evolve the lattice gauge variables forward in (conformal) time in parallel with the scalar field.
We will proceed by varying the action, \cref{fullAction}, with respect to the coefficients $U_\mu^a$, which again are the dynamical quantities of our system.
Since the plaquette action \cref{plaqAction} includes products of link variables at different lattice sites, we must be careful to vary the \textit{full} action,
\begin{align}
	S = \sum_{x,\tau} d\tau dx^3 \mathcal{L},
\end{align}
rather than the Lagrangian density $\mathcal{L}$.
That is, when varying with respect to the link $U_\gamma$ we must include all plaquettes at all lattice sites which include $U_\gamma$, which gives
\begin{align}\label{staplevariation}
	\begin{split}
		\frac{\partial S_G }{\partial U_\gamma^f(x)} &= \sum_{\mu \neq \gamma} \beta_{\gamma\mu} \Theta(x) \Tr{ \sigma^f W_{\gamma,\mu}(x) } \\
		&\hphantom{={} } + \sum_{\mu \neq \gamma} \beta_{\gamma\mu} \Theta(x-\hat{\mu}) \Tr{ \sigma^{f} W_{\gamma,-\mu}(x) }.
	\end{split}
\end{align}
In \cref{staplevariation} we encounter the \textit{staple} $W_{\gamma,\mu}(x)$, which is defined as the product of three links closing a plaquette about $U_\gamma(x)$,
\begin{align}
	U_\gamma(x) W_{\gamma,\mu}(x) = P_{\gamma\mu}(x),
\end{align}
the form one would expect to find upon differentiating a plaquette by one of its constituent links.

Before obtaining the explicit update rules for the link coefficients $U_\gamma^a$, we must first ensure that the unitarity constraint, \cref{unitarity}, is satisfied by the evolution equations.
To do so, we introduce to our theory a Lagrange multiplier term~\cite{GarciaBellido:2003wd},
\begin{align}
	\mathcal{L}_\lambda &= \sum_\mu \lambda_\mu(x) \Tr{ U_\mu(x) U_\mu^\dag(x) - 1 }.
	\label{lagmult}
\end{align}
The relevant variations are
\begin{align}
	\frac{\partial S_\lambda }{\partial U_\gamma^0(x)}
	= 4 \lambda_\gamma(x) U_\gamma^0(x)
\end{align}
and
\begin{align}
	\frac{\partial S_\lambda }{\partial U_\gamma^a(x)}
	= \lambda_\gamma(x) U_\gamma^a(x).
\end{align}
The resultant equation of motion (in matrix form) is
\begin{align}
	\begin{split}
		\lambda_\gamma(x) U_\gamma^\dag(x)
		&= \sum_{\mu\neq\gamma} \beta_{\gamma\mu} \Theta(x) W_{\gamma,\mu}(x) \\
		&\hphantom{={} } + \sum_{\mu\neq\gamma} \beta_{\gamma\mu} \Theta(x-\hat{\mu}) W_{\gamma,-\mu}(x).
	\end{split}
\end{align}
However, we seek to eliminate these Lagrange multipliers from our system of equations.
To do so, we first left-multiply both sides by the link $U_\gamma(x)$, obtaining
\begin{align}
\begin{split}
	\lambda_\gamma(x)
	&= \sum_{\mu\neq\gamma} \beta_{\gamma\mu} \Big( \Theta(x) P_{\gamma\mu}(x) + \Theta(x-\hat{\mu}) P_{\gamma(-\mu)}(x) \Big).
\end{split}
\end{align}
Note that
\begin{align}
	P_{\gamma(-\mu)}(x) &= U_\gamma(x) U_{-\mu}(x+\hat{\gamma}) U_i^\dag(x-\hat{\mu}) U_{-\mu}^\dag(x) \\
	&= U_\gamma(x) U_\mu^\dag(x+\hat{\gamma}-\hat{\mu}) U_\gamma^\dag(x-\hat{\mu}) U_\mu(x-\hat{\mu}).
\end{align}
Next, taking the trace-free part of this equation eliminates the Lagrange multiplier $\lambda_\gamma(x)$ and provides a system of equations for the three ``Pauli flavors'' $U_\gamma^a$ for $a=1$, 2, and 3, which is
\begin{align}\label{linkeomA}
	0 &= \sum_{\mu\neq\gamma} \beta_{\gamma\mu} \Big( \Theta(x) P_{\gamma\mu}^a(x) + \Theta(x-\hat{\mu}) P_{\gamma(-\mu)}^a(x) \Big).
\end{align}
\Cref{linkeomA} alone does not provide an explicit instruction on obtaining links at the next time step $U_\gamma(x + \hat{0})$ from the current links $U_\gamma(x)$.
However, \cref{linkeomA} does include (products of) links at the current and next time slices with the plaquettes $P_{0\gamma}$.
After fixing a temporal gauge, setting $U_0(x)=1$, we define these temporal plaquettes as a new dynamical field in our system, the \textit{electric field}
\begin{align}
	E_i(x) &= U_i(x + \hat{0}) U_i^\dag(x) = P_{0i}(x).
\end{align}
The temporal gauge fixes $E_0 = 1$.
This definition immediately shows that we may evolve the links by performing the matrix multiplication
\begin{align}\label{linkupdate}
	U_i(x + \hat{0}) &= E_i(x) U_i(x).
\end{align}
Namely, \cref{linkupdate} is the update rule for the links fields which gives the exact value for the $SU(2)$ matrix $U_i(x)$ at the next time slice, $\tau+\hat{0}$, in terms of the electric and links fields at the present time step $\tau$.
Note that while $E_i$ is not the physical electric field reminiscent of electromagnetism, it gets its name for its use in computing the electric field strength $\vert E \vert^2$.
The electric field component of the total energy density of the (coupled) gauge field takes the form
\begin{align}
	\rho_E
	&= \frac{1}{a(t)^4} \frac{2 \Theta(x)}{\kappa^2 a_i^4 g^2} \text{Tr}\left[ 1 - E_i \right] \\
	&= \frac{1}{a(t)^4} \frac{2 \Theta(x)}{\kappa^2 a_i^4 g^2} \cdot 2 \left( 1 - E_i^0 \right).
\end{align}

Now, the effect of including a Lagrange multiplier, \cref{lagmult}, and subsequently taking the trace-free part of the equation of motion, \cref{linkeomA},
is that we have update rules for only $E_i^a$, $a \neq 0$.
These are obtained by substituting the definition of the electric field into the full equation of motion, \cref{linkeomA}, which leads to
\begin{align}\label{eupdate}
\begin{split}
	E_i^a(x + \hat{0}) &= \frac{\Theta(x)}{\Theta(x + \hat{0})} E_i^a(x)
	- \kappa^2 \sum_{j\neq i} P_{ij}^a(x + \hat{0}) \\
	&\hphantom{={}} + \kappa^2 \sum_{j\neq i} \frac{\Theta(x + \hat{0} - \hat{\jmath})}{\Theta(x + \hat{0})} P_{ij}^a(x + \hat{0} - \hat{\jmath}).
\end{split}
\end{align}
With \cref{linkupdate} and \cref{eupdate}, we evolve the fields and their ``derivatives'' as one normally would when numerically integrating a second order differential equation.
The ``zero flavor'' of the electric field, $E_i^0$, which has no update rule, is instead computed via the unitarity constraint [identical to that for the links, \cref{unitarity}].\footnote{
The numerical computation of \cref{unitarity} via
\begin{align}
	E_\mu^0 &= \sqrt{1 - \frac{1}{4} \left( E_\mu^a \right)^2 },
\end{align}
only produces numerically significant values to $\sim 10^{-15}$, the threshold of double precision.
The initial fields are often small enough that $1 - E_\mu^0$, using the above computation, would be zero to machine precision.
Computing instead
\begin{align}
	1 - E_\mu^0
	= \frac{\frac{1}{4} \left( E_\mu^a \right)^2}{1 + \sqrt{1 - \frac{1}{4} \left( E_\mu^a \right)^2}}
\end{align}
provides full numerical precision for any size $E_i^0$.}
Finally, because the update rules produce exact computations for the field configurations from time slice to time slice, the gauge condition is preserved trivially, as is Gauss's law~\cite{Ambjorn:1990pu,Ambjorn:1990wn,Moore:1996wn,Moore:1997sn,Rajantie:2000nj,Tranberg:2003gi,GarciaBellido:2003wd}.


\section{Gauss's law and relaxation}\label{Arelax}

We now discuss the treatment of Gauss's law in our simulations.
As is familiar from electromagnetism, Gauss's law generally takes the form
\begin{align}
	D_i E_i = \rho,
\end{align}
where $D_i$ is the appropriate covariant derivative and $\rho$ is the charge density sourced by fields covariantly coupled to (i.e., charged under) the gauge field of interest.
($E_i = F_{0i}$ here represents the physical electric field, not the construct from lattice gauge theory.)
For the electromagnetic field [$U(1)$], the derivative $D_i$ is merely a spatial gradient $\partial_i$ as the photon is not charged under its own gauge group.
In the temporal gauge, Gauss's law may be imposed exactly in momentum space by setting the divergence of $A_i'$ equal to the appropriate charge density, i.e.,
\begin{align}
	A_i'(k) = \frac{ k_i }{k^2} \rho(k).
\end{align}

In contrast, the couplings both to the scalar and between non-Abelian fields make Gauss's law for our model nonlinear.
First, non-Abelian fields are charged under the adjoint representation of their own gauge group, which modifies the covariant derivative to
\begin{align}
	D_i A_j^a{}' \equiv \partial_i A_j^a{}' + f^{abc} A_i^b A_j^c{}'.
\end{align}
Next, we can think of the effect of the conformal coupling $\Theta(\phi) F^2$ as a further modification of the gauge-covariant derivative, which would read
\begin{align}
	D_i A_j^a{}' \equiv \partial_i A_j^a{}' + \frac{d \Theta/ d \phi}{\Theta} \partial_i \phi A_j^a{}' + f^{abc} A_i^b A_j^c{}'.
\end{align}
Here it is apparent that the scalar coupling makes Gauss's law an \textit{implicit} constraint on the $A_i^a{}'$.
That is, not only do the scalar and the other gauge fields $A_i^b$ and $A_i^b{}'$ for $b \neq a$ source the spatial divergence of $A_i^a{}'$, but $A_i^a{}'$ also sources its own divergence.

In the remainder of this Appendix we detail the implementation of the relaxation method described in \cref{sec:inits}.
Recall that the equation of motion for the temporal links $U_0(x)$ is Gauss's law for the lattice variables.
The local Gauss constraints $G^a(x)$ ($a \in \{1,2,3\}$) are defined by
\begin{align}
		G^a &\equiv \sum_{i\neq 0} \left( \Theta(x) E_i^a(x) - \Theta(x-\hat{\imath}) \tilde{E}_i^a(x) \right),
\end{align}
with $\tilde{E}$ the parallel-transported electric field
\begin{align}
		\tilde{E}_i^a \equiv \text{Tr}\left[ - 2 \sigma^a U_i^\dag(x-\hat{\imath}) E_i(x-\hat{\imath}) U_i(x-\hat{\imath}) \right].
\end{align}
As a measure of the satisfaction of Gauss's law, $G^a(x) = 0$, we define the ``Hamiltonian''
\begin{align}
	H = \sum_x G^a(x) G^a(x).
\end{align}
We seek to minimize $H$ by an iterative method, incrementing the electric fields $E_i^a$ in the direction which decreases $H$.
To do so, we evolve the dissipative equations~\cite{Rajantie:2000nj}
\begin{align}
	\frac{\partial E_j^c(x)}{\partial t} = - \frac{\partial H}{\partial E_j^c(x)}
\end{align}
through the fictitious time $t$, which increments the fields in the direction opposite to the gradient of $H$ in the space of the electric field variables.
Evaluating the derivative of $H$,
\begin{align}
	\begin{split}
		\frac{\partial H}{\partial E_i^a(x)}
		&= 2 \Theta(x) G^a(x) \\
		&\hphantom{ {}= } - 2 \Theta(x) G^c(x+\hat{\imath}) \text{Tr}\left[ - 2 \sigma^c U_i^\dag(x) \sigma^a U_i(x) \right].
	\end{split}
\end{align}
Note that this variation with respect to $E_j^a(x)$ picks up a contribution from both Gauss's law at $x$ and at $x+\hat{\jmath}$, since the relaxation must ensure that the global, grid-summed Gauss violation is improved, rather than independently improving each local Gauss violation.
Not only are the various $E_i^a$ coupled to each other, but discretized derivatives (or difference operators) directly couple fields at $x$ to the fields at lattice sites neighboring $x$.

Numerically, each step of the relaxation consists of incrementing each electric field by an amount proportional to the above derivative by a fictitious time step $\Delta t$,
\begin{align}
	\Delta E_i^a(x) = - \frac{\partial H}{\partial E_i^a(x)} \Delta t.
\end{align}
We track both the grid-averaged Gauss violation and electric field energy density over the course of relaxation, iterating until the grid-averaged Gauss violation drops below a preset tolerance.

Since the initial field configuration of electric fields (and link fields) is well approximated to linear order, the initial Gauss constraints $G^a(x)$ are proportional to $g$.
Therefore, we choose the tolerance to be proportional to $g$, relaxing the Gauss violation to a fixed fraction of the initial degree of violation.
In the results presented above, the relaxed Gauss violation is never larger than $10^{-5}$ times the initialized value.
Further, this choice ensures that the method requires an identical number of steps for any value of $g$, and that all relaxed configurations contain the same electric field energy density.

Finally, the choice of $\Delta t$ must be chosen to achieve convergence in the same sense as any other numerical integration: $\Delta t$ must be small enough that further reducing the time step has no effect on the resulting final field configuration.
We find that scaling $\Delta t$ by $\Theta(x)^{-1}$ [i.e., setting $\Delta t = \alpha / \Theta(x)$] ensures that the relaxed electric field energy density is also independent of the coupling parameter $M$.
Setting $\alpha = .1$ achieved satisfactory convergence for all parameters we tested.

\begin{figure}[htb]
\centering
\includegraphics[width=.99\columnwidth]{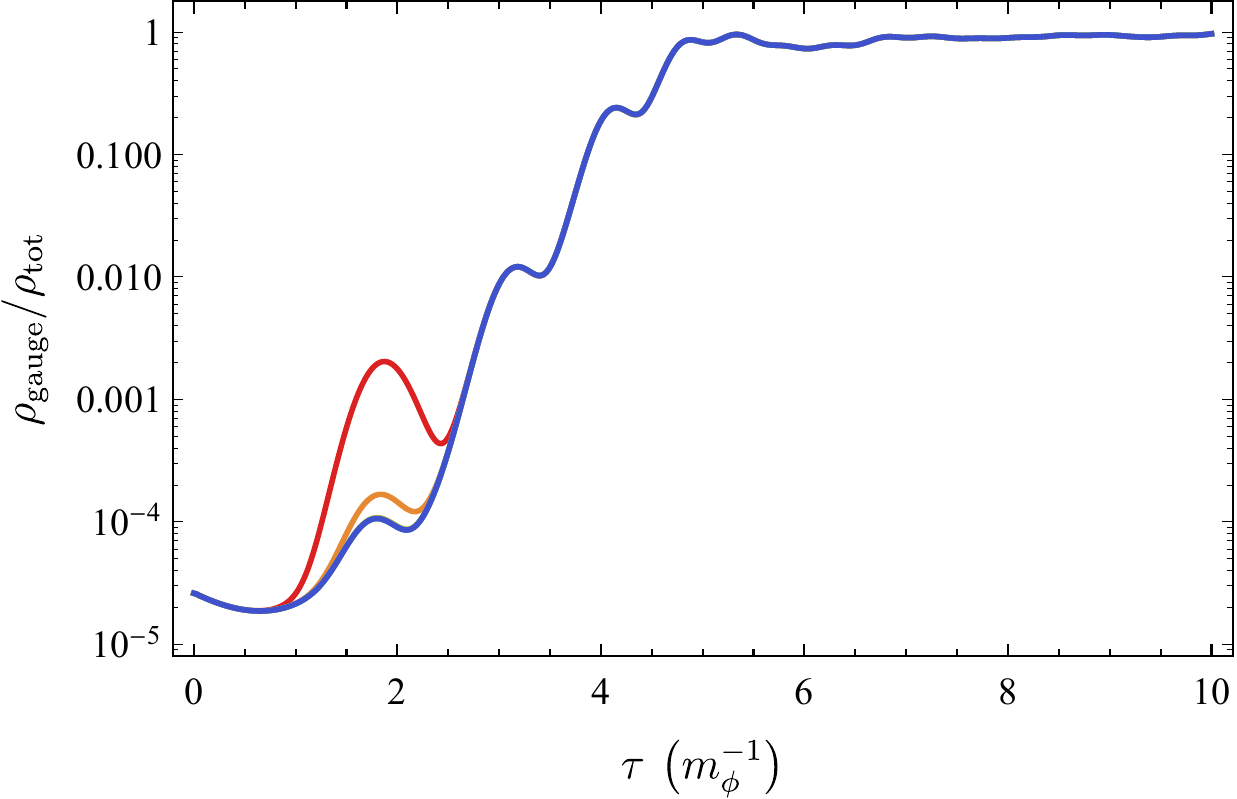} \\
\vspace{1eX}
\includegraphics[width=.99\columnwidth]{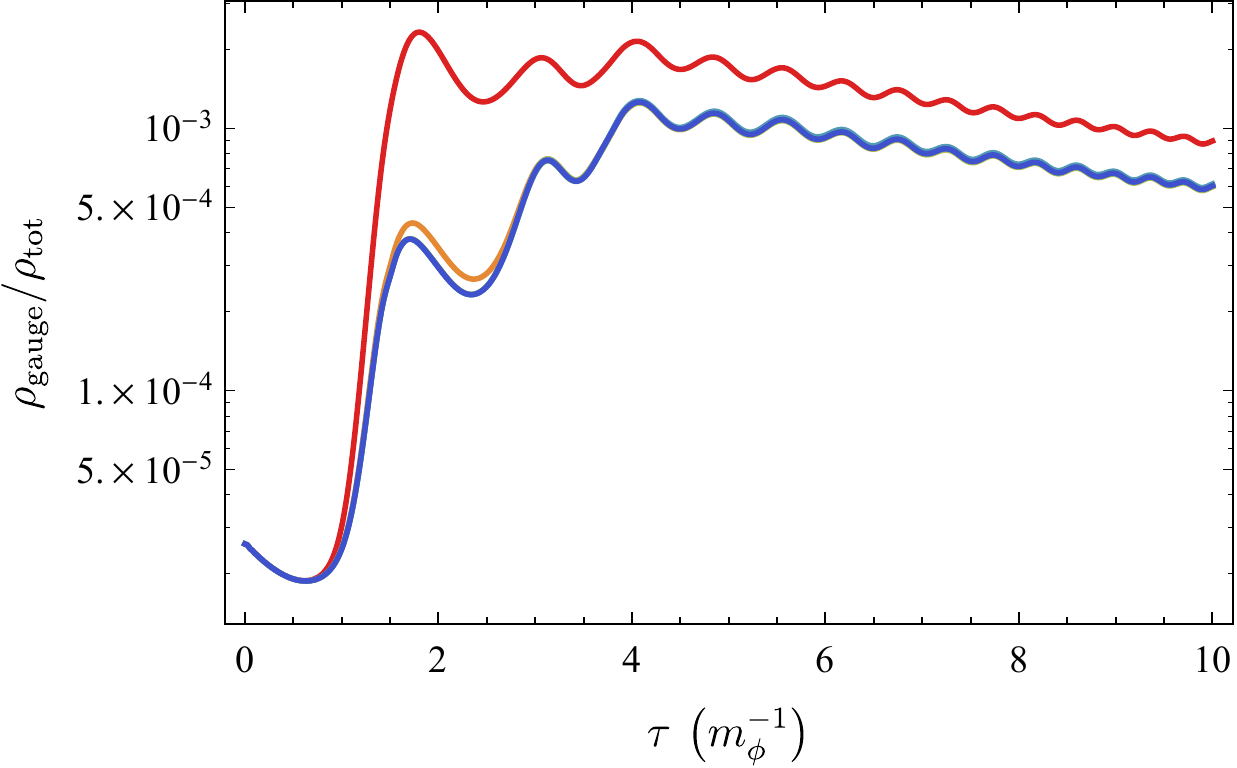}
\caption{
	The effect of increasing the degree of Gauss relaxation (from red to blue) for $M = .016 m_\text{pl}$, $g = 10^{-6}$ (top) and $M = .014 m_\text{pl}$, $g = 10^{-2}$ (bottom) on the evolution of $\rho_\text{gauge} / \rho_\text{tot}$.
	The colors red through blue correspond to relaxation of the grid-averaged Gauss violation $\abs{G^a}$ to $\sim 2$, 3, 4, 5, 6, and 7 orders of magnitude smaller than the initial value.
}\label{fig:moreRelax}
\end{figure}

\Cref{fig:moreRelax} demonstrates that, for two sets of couplings $g$ and $M$, the evolutions exhibit rapid convergence with respect to the amount of applied Gauss relaxation.
We use this metric as a guide for all simulations in this work, seeking to minimize computation time by relaxing only so far as has an appreciable effect on the evolution.

As expected, the effect of the violation of Gauss's law is a parametrically increasing artificial growth in gauge field strength~\cite{Moore:1996qs}.
In the case of weaker non-Abelian coupling, this is only evident in the resonance during the first oscillation, having minimal effect on the final configuration.
However, increasing the non-Abelian coupling strength has a more dramatic effect on the final energy fraction.


\section{The lattice action}\label{plaqToFAppendix}

In this appendix, we briefly outline the steps to verify the correspondence between plaquettes and $F^2$ given by \cref{plaqToF},
\begin{align}\label{plaqToFderivationAppendix}
	\text{Tr}\left[ \left( F_{\mu\nu}^a(x) \right)^2 \right] = \frac{4}{a_\mu^2 a_\nu^2 g^2} \text{Tr}\left[ 1 - P_{\mu\nu}(x) \right].
\end{align}
This identity may be verified by repeated application of the Baker-Campbell-Hausdorff formula,
\begin{align}
	\exp A \exp B = \exp\left( A + B + \frac{1}{2} \left[ A, B \right] + \cdots \right),
\end{align}
or by expanding the links in the limit of slowly varying gauge fields.
Here we outline the latter method, which begins by expanding the link definition, \cref{linkDefinition}, to second order in the lattice spacing, i.e.,
\begin{align}\label{linkapprox}
	U_\mu(x) \approx 1 - a_\mu g A_\mu^a(x) \sigma^a + \frac{1}{2} a_\mu^2 g^2 A_\mu^a(x) \sigma^a A_\mu^b(x) \sigma^b.
\end{align}
Plugging in this approximation to each of the four links within a plaquette,
\begin{align}\label{plaquetteforActionAppendix}
	P_{\mu\nu}(x) = U_\mu(x) U_\nu(x + \hat{\mu}) U_\mu^\dag(x + \hat{\nu}) U_\nu^\dag(x),
\end{align}
the $\mathcal{O}(a_\mu)$ terms of \cref{linkapprox} recover the Abelian component,
\begin{align}
	P_{\mu\nu}(x) = 1 - a_\mu a_\nu g \left( \partial_\mu A_\nu^a - \partial_\nu A_\mu^a \right) \sigma^a + \mathcal{O}(a_\mu^2),
\end{align}
where derivatives on the lattice are approximated by
\begin{align}
	\partial_\mu A_\nu^a(x) &= \frac{A_\nu^a(x+\hat{\mu})-A_\nu^a(x)}{a_\mu}.
\end{align}

The non-Abelian term is obtained by including $\mathcal{O}(a_\mu^2)$ terms in \cref{linkapprox}, taking here the fields to be slowly-varying between lattice sites.
This approximation cancels all terms with two copies of the same gauge field component, i.e., those resembling $A_\mu A_\mu$ or $A_\nu A_\nu$.
We are left with the desired non-Abelian terms of the form $A_\mu A_\nu$, such that
\begin{align}
	\begin{split}
		P_{\mu\nu}(x)
		&= 1 - a_\mu a_\nu g \left( \partial_\mu A_\nu^a - \partial_\nu A_\mu^a \right) \sigma^a \\
		&\hphantom{{}= } - a_\mu a_\nu g^2 \epsilon^{abc} A_\mu^b A_\nu^c \sigma^a + \mathcal{O}(a_\mu^4)
	\end{split} \\
	&= 1 - a_\mu a_\nu g \left( F_{\mu\nu}^a \sigma^a \right) + \mathcal{O}(a_\mu^4).
\end{align}

This term is still insufficient, as the trace of $1 - P_{\mu\nu}$ would be identically zero.
However, because the plaquette is unitary (or, if we take a hint from the Baker-Campbell-Hausdorff formula), we may infer that its exponential expansion must take the form
\begin{align}
	\begin{split}
		P_{\mu\nu}(x)
		&= 1 - a_\mu a_\nu g F_{\mu\nu}^a(x) \sigma^a \\
		&\hphantom{={} } + \frac{a_\mu^2 a_\nu^2 g^2}{2} F_{\mu\nu}^a(x) \sigma^a F_{\mu\nu}^b(x) \sigma^b + \mathcal{O}(a_\mu^6),
	\end{split}
\end{align}
which indeed satisfies \cref{plaqToFderivationAppendix}.


\bibliographystyle{apsrev4-1} 
\bibliography{SU2}

\end{document}